\documentclass[twocolumn,showpacs,preprintnumbers,amsmath,amssymb]{revtex4}
\usepackage{graphicx}
\usepackage{dcolumn}
\usepackage{bm}

\begin{document}

\title{Decoherence during Inflation: the generation
of classical inhomogeneities}

\author{Fernando C. Lombardo \footnote{lombardo@df.uba.ar}}
\author{Diana L\'opez Nacir \footnote{dnacir@df.uba.ar}}
\affiliation{Departamento de F\'\i sica {\it Juan Jos\'e
Giambiagi}, FCEyN UBA, Facultad de Ciencias Exactas y Naturales,
Ciudad Universitaria, Pabell\' on I, 1428 Buenos Aires, Argentina}

\date{\today}

\begin{abstract}
We show how the quantum to classical transition of the
cosmological fluctuations produced during inflation can be
described by means of the influence functional and the master
equation. We split the inflaton field into the system-field
(long-wavelength modes), and the environment, represented by its
own short-wavelength modes. We compute the decoherence times for
the system-field modes and compare them with the other time scales
of the model. We present the renormalized
stochastic Langevin equation for an homogeneous system-field and
then we analyze the influence of the environment on the power
spectrum for some modes in the system.
\end{abstract}


\maketitle

\section{Introduction}
The emergence of classical physics from quantum behaviour is
important for several physical phenomena in the early Universe.
This is beyond the fundamental requirement that only after the
Planck time can the metric of the Universe be assumed to be
classical. For example, the  inflationary era is assumed to have
been induced by scalar inflaton fields, with simple potentials
\cite{linde,Boya2004}. Such fields are typically assumed to have
classical behaviour, although in principle a full quantum
description should be used. In fact, the origin of large scale
structure in the Universe can be traced back to quantum
fluctuations that, after crossing the Hubble radius, were frozen
and became classical, stochastic, inhomogeneities \cite{todosLSS}.

It is generally assumed that several phase transitions have
occurred during the expansion of the Universe \cite{old}. As in
the case for the inflaton fields, the (scalar) order parameter
fields that describe these transitions are described classically.
However, the description of early universe phase transitions from
first principles is intrinsically quantum mechanical
\cite{cormier}. As a specific application \cite{Kibble} of the
previous point, the very notion of topological defects (e.g.
strings and monopoles) that characterize the domain structure
after a finite-time transition, and whose presence has
consequences for the early universe, is based on this assumption
of classical behaviour for the order parameter \cite{vilen}, as it
distributes itself between the several degenerate ground states of
the ordered system.

In previous publications, one of us has analyzed the emergence of a classical
order parameter during a second order phase transition and the
role of decoherence in the process of topological defect formation
\cite{lomplb,deconpb,diana,lomplb2}.

In the present paper our concern is directly related with the
first point above, the quantum to classical transition of the
inflaton. Any approach must take into account both the quantum
nature of the scalar field and the non-equilibrium aspects of the
process \cite{calzettahu95}. The problem of the quantum to
classical transition in the context of inflationary models was
first addressed by Guth and Pi \cite{guthpi}. In that work, the
authors used an inverted harmonic oscillator as a toy model to
describe the early time evolution of the inflaton, starting from a
Gaussian quantum state centered on the maximum of the potential.
They subsequently showed that, according to Schr\"odinger's
equation, the initial wave packet maintains its Gaussian shape
(due to the linearity of the model). Since the wave function is
Gaussian, the Wigner function is positive for all times. Moreover,
it peaks on the classical trajectories in phase space as the wave
function spreads. The Wigner function can then be interpreted as a
classical probability distribution for coordinates and momenta,
showing sharp classical correlations at long times. In other
words, the initial Gaussian state becomes highly squeezed and
indistinguishable from a classical stochastic process. In this
sense, one recovers a classical evolution of the inflaton rolling
down the hill.

A similar approach has been used by many authors to describe the
appearance of classical inhomogeneities from quantum fluctuations
in the inflationary era \cite{staro, Mopntemayor}. Indeed, a massless
free field $\phi$ in an expanding universe can be written as $\phi=a^{-1}\psi$
 where $a$ is the scale factor and the Fourier modes of the field $\psi$
satisfy the linear
equation
\begin{equation}
\psi_k''+(k^2-\frac{a''}{a})\psi_k = 0.
\end{equation}
For sufficiently long-wavelengths ($k^2\ll a''/a$), this equation
describes an unstable oscillator. If one considers an initial
Gaussian wave function, it will remain Gaussian for all times, and
it will spread with time. As with the toy model of Guth and Pi,
one can show that classical correlations do appear, and that the
Wigner function can again be interpreted as a classical
probability distribution in phase space. (It is interesting to
note that a similar mechanism can be invoked to explain the origin
of a classical, cosmological magnetic field from amplification of
quantum fluctuations).

However, classical correlations are only one aspect of classical
behaviour. It was subsequently recognized that, in order to have a
complete classical limit, the role of the environment is crucial,
since its interaction with the system distinguishes the field
basis as the pointer basis \cite{kiefer}. [We are reminded that,
even for the fundamental problem of the space-time metric becoming
classical, simple arguments based on minisuperspace models suggest
that the classical treatment is only correct because of the
interaction of the metric with other quantum degrees of freedom
\cite{halli}.]

While these linear instabilities cited above characterize   {\it
free} fields, the approach fails when interactions are taken into
account. Indeed, as shown again in simple quantum mechanical
models (e.g. the anharmonic inverted oscillator), an initially
Gaussian wave function becomes non-Gaussian when evolved
numerically with the Schr\"odinger equation. The Wigner function
now develops negative parts, and its interpretation as a classical
probability breaks down \cite{diana}. One can always force the
Gaussianity of the wave function by using a Gaussian variational
wave function as an approximate solution of the Schr\"odinger
equation, but this approximation deviates significantly from the
exact solution as the wave function probes the non-linearities of
the potential \cite{diana,stancioff}.

When interactions are taken into account, classical behaviour is
recovered only for ``open systems'', in which the observable
degrees of freedom interact with their environment. When this
interaction produces {\it both} a diagonalization of the reduced
density matrix and a positive Wigner function, the quantum to
classical transition is completed \cite{giulinibook}.

In Ref. \cite{diana} has been studied an anharmonic inverted
oscillator coupled to a high temperature environment. Under some
considerations, it was shown that the system becomes classical
very quickly, even before the wave function probes the
non-linearities of the potential.  Being an early time event, the
quantum to classical transition can now be studied perturbatively.
In general, recoherence effects are not expected \cite{nuno}.
Taking these facts into account, we have extended the approach to
field theory models \cite{lomplb,deconpb}. In field theory, one is
usually interested in the long-wavelengths of the order parameter.
Even the early universe is replete with fields of all sorts which
comprise a rich environment, in the inflationary example, we
considered a model in which the system-field interacts with the
environment-field, including only its own short-wavelengths. This
is enough during inflation. Assuming weak self-coupling constant
(a flat inflaton potential) we have shown that decoherence is an
event shorter than the time $t_{\rm end}$, which is a typical
time-scale for the duration of inflation. As a result,
perturbative calculations are justified \cite{deconpb}. Subsequent
dynamics can be described by a stochastic Langevin equation, the
details of which are only known for early times \cite{matacz}.

In our approach, the quantum to classical transition is defined by
the diagonalization of the reduced density matrix. In phase
transitions the separation between long and short-wavelengths is
determined by their stability, which depends on  the parameters of
the potential. During Inflation, this separation is set by the
existence of the Hubble radius. Modes cross the apparent horizon
during their evolution, and they are usually treated as classical.
The main motivation of this paper is to present a formal way to
understand this statement within the open quantum system approach.
In the last sense, decoherence is the critical ingredient if we
are to dynamically demonstrate the quantum-to-classical transition
of the open system.

The splitting between short and long-wavelength modes may be done using
a time-dependent or time-independent comoving cut-off as well as a
smoother window function \cite{matacz,Riotto}. Since
any time-dependent splitting produce an effective and arbitrary (split-dependent)
interaction between the system and environment degrees of freedom,
which can only be discarded by making some additional assumptions,
we consider convenient to use a time-independent comoving
cut-off, as it has been done in Refs. \cite{HUPAZYHANG,lombmazz}

The paper is organized as follows. In Section II we introduce our
model. This is a theory containing a real system-field, massless
and minimally coupled to a fixed de Sitter background. We compute
the influence functional by integrating out the environmental
sector of the field, composed by the short-wavelength modes.
Section III is dedicated to reviewing the evaluation of the master
equation and the diffusion coefficients which are relevant in
order to study decoherence. In Section IV we analyze the diffusion
coefficients and evaluate upper bounds on the decoherence times.
As we will see, decoherence takes place before the end of the
inflationary period. Section V is concerned with the effective
stochastic evolution of the system. We present the renormalized
stochastic Langevin equation for an homogeneous system-field and
then we analyze the influence of the environment on the power
spectrum for some modes in the system. Section VI contains our
final remarks. Two short appendices fill in some of the detail.
Throughout the paper we use units such that $\hbar=c=1$.

\section{The Influence functional and the density matrix}

Let us consider a massless quantum scalar field, minimally coupled
to a de Sitter spacetime $ds^2 = a(\eta ) [d\eta^2 - d\vec x^2]$
(where $\eta$ is the conformal time, $d\eta = dt/a(t)$ with  $t$
the cosmic time), with a quartic self-interaction. The classical
action is given by
\begin{equation}\label{desitteraction}
S[\phi] = \int d^4x ~a^4(\eta )~\left[\frac{{\phi'}^2}{2a^2(\eta
)} - \frac{{\nabla\phi}^2}{2a^2(\eta )} - \lambda \phi^4\right],
\end{equation}
where $a(\eta) = -1/(H\eta)$ and $\phi'=d\phi/d\eta$ ($a(\eta_i) =
1$ [$\eta_i = - H^{-1}$], and $H^{-1}$ is the Hubble radius). Let us
make a system-environment field splitting \cite{lombmazz}

\begin{equation}
\phi = \phi_< + \phi_>,
\end{equation}
where the system-field $\phi_<$ contains the modes with wave vectors shorter
than a critical value $\Lambda\equiv{}2\pi/\lambda_c$, while the
environment-field $\phi_>$ contains wave vectors longer than $\Lambda$. As we
set $a(\eta_i) = 1$, a physical length $\lambda_{\rm phys} =
a(\eta)\lambda$ coincides with the corresponding comoving length
$\lambda$ at the initial time. Therefore, the splitting between system and
environment gives a system sector constituted by all the modes
with physical wavelengths shorter than the critical length
$\lambda_c$ at the initial time $\eta_i$.

After splitting, the total action (\ref{desitteraction}) can be
written as
\begin{equation}S[\phi] = S_{0}^{<}[\phi_<] + S_{0}^{>}[\phi_>] + S_{\rm int}[\phi_<,
\phi_>],\label{actions}\end{equation} where $S_0$ denotes the free
field action and the interaction terms are given by
\begin{eqnarray}&& S_{\rm int}[\phi_<, \phi_>] = - \lambda \int d^4x ~a^4(\eta)~
\left\{\phi_<^4(x) + \phi_>^4(x) \right. \nonumber \\
&+& \left. 6 \phi_<^2(x) \phi_> ^2(x) + 4 \phi_<^3(x) \phi_>(x) +
4 \phi_<(x) \phi_>^3(x)\right\}.\label{inter}\end{eqnarray}

The total density matrix elements (for the system and environment fields)
are defined as
\begin{equation}
 \rho[\phi^{+}_{<},\phi^{+}_{>}|\phi^{-}_{<},\phi^{-}_{>};\eta]=\langle\phi^{+}_{<}\phi^{+}_{>}|
 \hat{\rho}[\eta]|\phi^{-}_{<}\phi^{-}_{>}\rangle ,\label{matrix}
\end{equation}
where $\vert \phi_<^\pm\rangle$ and $\vert \phi_>^\pm \rangle$ are
the eigenstates of the field operators ${\hat\phi}_<$ and
${\hat\phi}_>$, respectively. For simplicity, we will assume that
the interaction is turned on at the initial time $\eta_i$ and
that, at this time, the system and the environment are not
correlated (we ignore, for the moment, the physical consequences
of such a choice, it has been discussed in \cite{deconpb}).
Therefore, the total density operator can be written as the
product of the density  operator for the system and for the
environment
\begin{equation}\label{sincorr}    \hat{\rho}[\eta_{i}]=\hat{\rho}_{>}[\eta_{i}]\hat{\rho}_{<}[\eta_{i}].
\end{equation} We will further assume that the initial state of
the environment is the Bunch-Davies vacuum \cite{BirreLLDavies}.

We are interested in the influence of the environment on the
evolution of the system. Therefore the reduced density matrix is
the object of relevance. It is defined by
\begin{equation}\rho_{\rm r}[\phi^{+}_{<}|\phi^{-}_{<};\eta]=
\int\mathcal{D}\phi_{>}\rho[\phi^{+}_{<},\phi_{>}|\phi^{-}_{<},\phi_{>};\eta].\label{red}
\end{equation}
The reduced density matrix evolves in time by means of
\begin{eqnarray}\rho_{\rm r}[\phi^{+}_{<f}|\phi^{-}_{<f};\eta]&=&\int d\phi_{<i}^+\int
d\phi_{<i}^-\;\rho_{r}[\phi^{+}_{<i}|\phi^{-}_{<i};\eta_i]\\\nonumber
&\times& J_{\rm
r}[\phi^{+}_{<f},\phi^{-}_{<f};\eta|\phi^{+}_{<i},\phi^{-}_{<i};\eta_i],\label{evol}
\end{eqnarray}
where $J_{\rm r}$ is the reduced evolution operator
\begin{eqnarray}J_{\rm
r}[\eta|\eta_i]&=&
\int_{\phi_{<i}^+}^{\phi_{<f}^+}\mathcal{D}\phi_{<}^+
\int_{\phi_{<i}^-}^{\phi_{<f}^{-}}\mathcal{
D}\phi_{<}^{-} \;\;\;\;\;\;\;\;\;\;\;\;\;\;\;\;\;\;\;\;\;\;\;\;\;\;\;\nonumber \\
&\times& \exp\{i(S^{<}[\phi_<^+]-S^{<}[\phi_<^-])\}
F[\phi_<^+,\phi_<^-].\label{evolred}
\end{eqnarray}
The  influence functional (or Feynman-Vernon functional)
$F[\phi_<^+,\phi_<^-]$ is defined as
\begin{eqnarray}
&& F[\phi^+_<,\phi^-_<] = \int d\phi^+_{{> i}} \int d\phi^-_{{>i}}
~ \rho_{>} [\phi_{{>i}}^+,\phi_{{>i}}^-,\eta_i] \int
d\phi_{{>f}}  \nonumber\\
&& \times \int_{\phi^+_{{>i}}}^{\phi_{{>f}}}\mathcal{D}\phi^+_{>}
\int_{\phi^-_{{>i}}}^{\phi_{{>f}}} \mathcal{D}\phi^-_{>}
\exp\left\{i(S[\phi^+_{>} ]+S_{{\rm int}}
[\phi^+_<,\phi^+_{>}])\right\} \nonumber \\
&&\times  \exp\left\{-i(S[\phi^-_{>}] + S_{{\rm
int}}[\phi^-_<,\phi^-_{>}]) \right\}.
\end{eqnarray}
This functional takes into account the effect of the environment
on the system. The influence functional describes the averaged
effect of the environmental degrees of freedom on the system
degrees of freedom to which they are coupled. With this
functional, one can identify a noise and dissipation kernel
related by some kind of fluctuation-dissipation relation. This
relation is important when one is interested in possible
stationary states where a balance is eventually reached. During
inflation we have the field (inflaton) on a very flat potential, 
away from its
minimum, and we are, in general, only interested in the dynamics
over some relatively small time. For example, we would neglect
dissipation during the slow-roll period; but it is not correct
during the eventual reheating phase.

We define the influence action $\delta A[\phi_<^+,\phi_<^-]$ and
the coarse grained effective action (CGEA) $A[\phi_<^+,\phi_<^-]$
as
\begin{equation}F[\phi_<^+,\phi_<^-] = \exp\{i
\delta A[\phi_<^+,\phi_<^-]\},\label{IA}\end{equation}
\begin{equation}A[\phi_<^+,\phi_<^-] = S[\phi_<^+] - S[\phi_<^-] + \delta
A[\phi_<^+,\phi_<^-].\label{CTPEA}\end{equation} We will calculate
the influence action perturbatively in $\lambda$ and we will
consider only terms up to order $\lambda^2$ and one loop in the
$\hbar$ expansion. The influence action has the following form \cite{lombmazz}:
\begin{eqnarray}
\delta{A}[\phi^{+}_{<},\phi^{-}_{<}]&=&\langle{}S_{\rm int}[\phi^{+}_{>},\phi^{+}_{<}]\rangle_{0}
-\langle{}S_{\rm int}[\phi^{-}_{>},\phi^{-}_{<}]\rangle_{0}\\\nonumber
&-&i\langle{}S_{\rm int}[\phi^{+}_{>},\phi^{+}_{<}]S_{\rm int}[\phi^{-}_{>},\phi^{-}_{<}]
\rangle_{0}\\\nonumber
&+&i\langle{}S_{\rm int}[\phi^{+}_{>},\phi^{+}_{<}]\rangle_{0}\langle{}S_{\rm int}[\phi^{-}_{>},
\phi^{-}_{<}]\rangle_{0}\\
\nonumber
&+&\frac{i}{2}\left\{\langle{}S_{\rm int}[\phi^{+}_{>},\phi^{+}_{<}]\rangle^{2}_{0}-\langle{}
S_{\rm int}[\phi^{+}_{>},\phi^{+}_{<}]^{2}\rangle_{0}\right\}\\
\nonumber
&+&\frac{i}{2}\left\{\langle{}S_{\rm int}[\phi^{-}_{>},\phi^{-}_{<}]\rangle^{2}_{0}-\langle{}
S_{\rm int}[\phi^{-}_{>},\phi^{-}_{<}]^{2}\rangle_{0}\right\}
\end{eqnarray}
where $\langle ~\rangle_0$ is the quantum average with respect to
the free field action of the environment, defined as
\begin{eqnarray}\label{valormedio}
&&\langle{}B[\phi^{+}_{>},\phi^{-}_{>}]\rangle_{0}=\int{d}\phi^{+}_{>i}
\int{d}\phi^{-}_{>i}\int{d}\phi_{>f}\\
\nonumber&&\times\int_{\phi^{+}_{>i}}^{\phi_{>f}}{D}\phi^{+}_{>}
\int_{\phi^{-}_{>i}}^{\phi_{>f}}{D}\phi^{-}_{>}B[\phi^{+}_{>},\phi^{-}_{>}]\;\;\\
\nonumber
&&\times\exp\left\{i\left(S^{>}[\phi^{+}_{>}]-S^{>}[\phi^{-}_{>}]
\right)\right\}\langle\phi^{+}_{>i}|\hat{\rho}_{>}[\eta_{i}]|\phi^{-}_{>i}\rangle.
\end{eqnarray}
Here $\hat{\rho}_{>}$ is the Bunch-Davies vacuum state assumed for
the environment.

The influence functional can be computed, and the result is
\begin{eqnarray}\nonumber{\rm Re}\delta A&=&-\lambda\int d^4x_1\;a^4(\eta)
\left\{2\Delta_{4}(x_1)\right.\\\nonumber
&-&\left.12\Delta_{2}(x_1)i{G}^{\Lambda}_{++}(x_1,x_1)\right\}\\\nonumber
&+& \lambda^2 \int d^4x_1\int d^4x_2 ~a^4(\eta_1)~a^4(\eta_2) ~ \Theta (\eta_1 - \eta_2) \nonumber \\
&\times& \left\{64 \Delta_3(x_1) {\rm Re}G^\Lambda_{++}(x_1,x_2)
\Sigma_3(x_2) \right. \nonumber \\
&+&\left. 288 \Delta_2(x_1) {\rm Im}G^{\Lambda 2}_{++}(x_1,x_2)
\Sigma_2(x_2)\right\},
\label{inff}\\
{\rm Im}\delta A&=&  \lambda^2 \int
d^4x_1 \int d^4x_2 ~a^4(\eta_1)~ a^4(\eta_2) \nonumber \\
&\times& \left\{ 32 \Delta_3(x_1) {\rm Im}G^\Lambda_{++}(x_1,x_2)
\Delta_3 (x_2)\right. \nonumber \\
&-& \left. 144 \Delta_2(x_1) {\rm Re}G^{\Lambda 2}_{++}(x_1,x_2)
\Delta_2 (x_2) \right\},\label{inffimag}
\end{eqnarray} where $x_j$ denotes
($\eta_j$,$\vec{x}_j$), $\Theta(x)$ is the Heaviside step function, and the integrations in time run from
$\eta_i$ to $\eta$.
$G^{\Lambda}_{++}(x_1,x_2)\equiv{}i\langle\phi_{>}^{+}(x_1)\phi_{>}^{+}(x_2)\rangle_0$
is the relevant short-wavelength closed time-path correlator
(it is proportional to the Feynmann propagator of the environment
field, where the integration over momenta is restricted by the
presence of the infrared cut-off $\Lambda$), and we have defined
\begin{equation}\Delta_n =\frac{1}{2}(\phi_<^{+n} - \phi^{-n}_<)\;\;\;
,\;\;\; \Sigma_n =\frac{1}{2}(\phi_<^{+n} + \phi_<^{-n}),
\end{equation}with $n=1,2,3$.

\section{Master equation and diffusion coefficients}

In this Section we obtain the evolution equation for the
reduced density matrix (master equation), paying particular
attention to the diffusion terms, which are responsible for
decoherence. To do so, we closely follow the quantum Brownian motion
(QBM) example \cite{unruh,qbm}, translated into quantum field
theory \cite{lomplb,lombmazz}.

The first step in the evaluation of the master equation is the
calculation of the density matrix propagator $J_{\rm r}$ from
Eq.(\ref{evolred}). In order to solve the functional integration
which defines the reduced propagator, we perform a saddle point
approximation, assuming the classical field configuration 
dominates functional integrals, 
\begin{equation}
J_{\rm
r}[\phi^+_{<f},\phi^-_{<f},\eta\vert\phi^+_{<i},\phi^-_{<i},
\eta_i] \approx \exp{ i A[\phi^+_{<\rm cl},\phi^-_{<\rm cl}]},
\label{prosadle}
\end{equation}
where $\phi^{\pm}_{<\rm cl}$ is the solution of the semiclassical
equation of motion ${\delta Re
A/\delta\phi_<^+}\vert_{\phi_<^+=\phi_<^-}=0$ with boundary
conditions $\phi^{\pm}_{<\rm cl}(\eta_i)=\phi^{\pm}_{<i}$ and
$\phi^{\pm}_{<\rm cl}(\eta)=\phi^\pm_{<f}$. Since we are working up
to $\lambda^2$ order, we can evaluate the influence functional
using the solutions of the free field equations. This classical
equation is $ \phi_<^{''} + 2 \mathcal{ H}\phi_<^{'} -
\nabla^2\phi_< = 0, $ ($\mathcal{ H} = a'(\eta)/a(\eta)$). A
Fourier mode $\psi_{\vec k}$ of the field $\psi\equiv a(\eta)\phi_{<}$,
satisfies
\begin{equation}
\psi_{\vec k}^{''} + \left(k^2 - \frac{2}{\eta^2}\right)\psi_{\vec k}
= 0, \label{newmodes}\end{equation} where we have used the fact that $a''/a= 2/\eta^2$.
It is important to note that for long-wavelength modes, $k \ll
2/\eta^2$, Eq. (\ref{newmodes}) describes an unstable
(upside-down) harmonic oscillator \cite{guthpi}.

The classical solution for a mode in the system can be written as
\begin{equation}
\phi_{\vec k}^{\pm \rm cl}(\eta') = \phi_{< i}^\pm(\vec k)u_1(\eta',
\eta) + \phi_{< f}^\pm(\vec k)u_2(\eta',
\eta),\label{classmode}\end{equation} where
\begin{subequations}
\begin{align}
u_1=&\frac{\sin[k(\eta-\eta')](\frac{1}{k}+k\eta\eta')+\cos[k(\eta -\eta')](\eta'-\eta)}{
\sin[k(\eta'-\eta_i)](\frac{1}{k}+k\eta_i\eta')+\cos[k(\eta'-\eta_i)](\eta_i-\eta')},\nonumber \\
u_2 =& \frac{\sin[k(\eta_i - \eta')](\frac{1}{k} +k\eta'\eta_i)+
\cos[k(\eta' - \eta_i)](\eta' - \eta_i)}{\sin[k(\eta_i - \eta)](\frac{1}{k} +k\eta\eta_i)+
\cos[k(\eta - \eta_i)](\eta - \eta_i)}.\nonumber
\end{align}\end{subequations}
We will assume that the system-field contains only one Fourier
mode with $\vec k = \vec k_0$. This is a sort of ``minisuperspace"
approximation for the system-field that will greatly simplify the
calculations, therefore we assume
\begin{equation}\label{classmodecomp}
\phi_{<\rm cl}^{\pm}(\vec x, \eta') = \phi_{\vec k_0}^{\pm \rm
cl}(\eta') \cos({\vec k}_0 .\vec x),
\end{equation}
where $\phi_{\vec k_0}^{\pm \rm cl}$ is given by
(\ref{classmode}).

In order to obtain the master equation we must compute the final
time derivative of the propagator $J_{\rm r}$. After that, all the
dependence on the initial field configurations $\phi^\pm_{<i}$
(coming from the classical solutions $\phi^{\pm\rm cl}_<$) must be
eliminated. Following the same procedure outlined in previous publications
\cite{deconpb}, we can
prove that the free propagator satisfies
\begin{eqnarray}\label{relacsoluc}
\phi^{\pm \rm cl}_{k_{0}}(\eta')J_{0}&=&\Big{\{}u_{2}(\eta',\eta)\phi^{\pm}_{< f}\\\nonumber
&-&\frac{2a^{2}(\eta)u'_{2}(\eta,\eta)u_{1}(\eta',\eta)\phi^{\pm}_{< f}}{a^{2}
(\eta)u'_{1}(\eta,\eta)-a^{2}(\eta_{i})u'_{2}(\eta_{i},\eta)}\\
\nonumber &\mp&\frac{4{
}i\;u_{1}(\eta',\eta)V^{-1}}{[a^{2}(\eta)u'_{1}(\eta,\eta)-a^{2}(\eta_{i})u'_{2}
(\eta_{i},\eta)]}\partial_{\phi^\pm_{\rm
< f}}\Big{\}}J_{0},
 \end{eqnarray}where a prime now stands for a derivative with
 respect to $\eta'$ and the spatial volume $V$ appears because of
 we are considering only one Fourier mode for the system. These identities
allow us to remove the initial field configurations
$\phi^\pm_{<i}$, by expressing them in terms of the final
amplitudes $\phi^\pm_{< f}$ and the derivatives
$\partial_{\phi^\pm_{< f}}$, and obtain the master equation.

The full equation is very complicated and, as for quantum Brownian
motion, it depends on the system-environment coupling. In what
follows we will compute the diffusion coefficients for the
different couplings described in the previous section. As we are
solely interested in decoherence, it is sufficient to calculate
the correction to the usual unitary evolution coming from the
imaginary part of the influence action. The result
reads
\begin{eqnarray}\nonumber
&i&\partial_{\eta}\rho_{r}[\phi^{+}_{<f}|\phi^{-}_{<f};\eta]=\langle\phi^{+}_{<f}|
[\hat{H}_{\rm ren},\hat{\rho}_{r}]|\phi^{-}_{<f}\rangle\\\nonumber
&-&i\left[\Gamma_{1}{D}_{1}(\vec{k_{0}},\eta,\Lambda)+\Gamma_{2}{D}_{2}(\vec{k_{0}},
\eta,\Lambda)\right]\rho_{r}[\phi^{+}_{<f}|\phi^{-}_{<f};\eta]\\
&+&...\;, \label{master}\end{eqnarray} where we have defined
$\Gamma_{1}=\frac{\lambda^{2}V}{H^{2}}({\phi^{+}_{<f}}^{3}-{\phi^{-}_{<f}}^{3})^{2}$
and
$\Gamma_{2}=\frac{\lambda^{2}V}{4}({\phi^{+}_{<f}}^{2}-{\phi^{-}_{<f}}^{2})^{2}$.
The ellipsis denotes other terms coming from the time derivative
that not contribute to the diffusive effects. This equation
contains time-dependent diffusion coefficients $D_j$. Up to one
loop, only $D_1$ and $D_2$ survive. Coefficient $D_1$ is related
to the interaction term $\phi_<^3 \phi_>$, while $D_2$ to
$\phi_<^2 \phi_>^2$. These coefficients can be (formally) written
as
\begin{eqnarray}\label{D1}
D_1(\vec{k_{0}},\eta,\Lambda)&=&\frac{H^{2}}{2}\int_{\eta_{i}}^{\eta}d\eta'a^{4}
(\eta)a^{4}(\eta')F^{3}_{\rm cl}(\eta,\eta',{k}_{0})\\
\nonumber&\times&{ImG}^{\Lambda}_{++}(\eta,\eta',3\vec{k_{0}})\;\Theta(3{k}_{0}-\Lambda),
\;\;\;\;\;\;\;\;\;\;\;\;\;
\end{eqnarray}
and
\begin{eqnarray}\label{D2}
{D}_{2}(\vec{k_{0}},\eta,\Lambda)&=&-36\int_{\eta_{i}}^{\eta}d\eta'a^{4}(\eta)
a^{4}(\eta')F^{2}_{\rm cl}(\eta,\eta',{k}_{0})\\
\nonumber &\times&
[{ReG}^{\Lambda 2}_{++}(\eta,\eta',2\vec{k_{0}})+2\;{ReG}^{\Lambda 2}_{++}(\eta,\eta',0)],
\end{eqnarray}
with the function $F_{\rm cl} $ defined by
\begin{equation}\label{Fcl} F_{\rm
cl}(\eta,\eta_{i},{k}_{0})=\frac{\sin[{k}_{0}(\eta-\eta_{i})]}{{k}_{0}\eta}+
\frac{\eta_{i}\cos[{k}_{0}(\eta-\eta_{i})]}{\eta}.
\end{equation}
The explicit expressions of these coefficients are complicated
functions of conformal time, the particular mode $k_0$, and the
cut-off $\Lambda$, and we show them in Appendix \ref{AppendixA}.

It is important to note that here we are only studying the effect of 
normal diffusion terms, even it is known that  
anomalous diffusion terms can also be relevant at zero temperature. Analysis done in Ref. \cite{qbm} 
suggests that anomalous diffusion for a supraohmnic environment is only relevant on a small  
transient and decoherence for unstable long-wavelength modes are driven by normal 
diffusion coefficients \cite{paula}. 
\section{Decoherence}

Coherences are destroyed by diffusion terms. This process is evident 
after considering the following appoximate solution to the 
master equation
\begin{eqnarray} &\rho_{\rm r}&[\phi^+_<, \phi^-_<; \eta] \approx
\rho^{\rm u}_{\rm r}[\phi^+_<, \phi^-_<; \eta] \nonumber \\
&&\times  \exp \left[-\sum_j \Gamma_{\rm j} \int_{\eta_i}^{\eta_f}
d\eta ~D_{\rm j}(k_0,\Lambda,\eta) \right],
\end{eqnarray} where
$\rho^{\rm u}_{\rm r}$ is the solution of the unitary part of the
master equation (i.e. without environment), and $\Gamma_j$
includes the coefficients in front each diffusion term in
Eq.(\ref{master}). The system will decohere when the non-diagonal
elements of the reduced density matrix are much smaller than the
diagonal ones.

The decoherence time-scale sets the time after which we have a
classical field configuration, and it can be defined as the
solution to
\begin{eqnarray}\label{deftd}
1&\approx&\sum_{j}\Gamma_{j}\int_{\eta_{i}}^{\eta_{d}}d\eta\;{D}_{j}(k_{0},\Lambda,\eta)\\
\nonumber&\gtrsim&\Gamma_{l}\int_{\eta_{i}}^{\eta_{d}}d{}\eta\;{D}_{l}(k_{0},\Lambda,\eta),
\end{eqnarray}where the inequality is valid for any particular $j=l$. That is, the
interactions with the environment have a cumulative effect on the
onset of classical behaviour, i.e. the inclusion of a further
interaction term reduces the decoherence time $\eta_{d}$.
Therefore, in order to find upper bounds to $\eta_{d}$, we define
the decoherence time $\eta_{d_{j}}$ coming from each diffusion
coefficient by
\begin{equation}\label{deftdj}
1\approx\Gamma_{j}\int_{\eta_{i}}^{\eta_{d_{j}}}d\eta\;{D}_{j}(k_{0},\Lambda,\eta),
\end{equation} with $j=1$,$2$.
\subsection{Diffusion terms: Numerical results and analytic approximations}

In this subsection we will analyze the behaviour of each diffusion
coefficient as a function of the Fourier mode $k_{0}$ (considered
for the system in Eq.(\ref{classmodecomp})) and the cut-off
$\Lambda$. We will also analyze the temporal evolution of the
coefficients and their integration in time for fixed values of
$k_0$ and $\Lambda$.  We will present simple analytical approximations
to the coefficients which can be used in Eq.(\ref{deftdj}) instead of
the full expressions to estimate the decoherence time-scale.

We define the dimensionless quantity $\mathcal{N}[k_{0},\eta]$
which is the number of e-foldings between the time $\eta_{k_{0}}$
when the mode $k_{0}$ crosses the Hubble radius (i.e.,
$|k_{0}\eta_{k_0}|=1$) and any time $\eta$ during inflation,
\begin{equation}\label{Neta}
\mathcal{N}[k_{0},\eta]\equiv{}\ln\left|\frac{\eta_{k_{0}}}{\eta}\right|=-\ln{|k_{0}\eta|}.
\end{equation}
This quantity has the special feature that its sign indicates
whether the mode is inside ($\mathcal{N}[k_{0},\eta]<0$) or
outside ($\mathcal{N}[k_{0},\eta]>0$) the Hubble radius.

Let us first consider the diffusion coefficient $D_1$, which comes
from the interaction term $\phi^{3}_{<}\phi_{>}$. Because of we
are considering only one Fourier mode for the system, with wave
vector $\vec{k_{0}}$, and the environment-field contains only
modes with $k > \Lambda$, this coefficient is different from zero
only if $\Lambda/3<k_{0}<\Lambda$ (i.e., $\phi_{<}$ is only
coupled with the $\vec{k}=3\vec{k_{0}}$ mode of the environment).

For this coefficient we can obtain an exact analytical expression
from Eq.(\ref{coefD1}) (see Appendix \ref{AppendixA}). In Fig.
\ref{d1kG} we have plotted this expression as a function of
$k_{0}$ for a particular value of the conformal time
($H\eta=-1/2$). For later times, the graphs are qualitatively
similar but $D_1$ oscillates more rapidly (since we have obtained
 our results by perturbative calculations, they are not valid at large
times).
The coefficient decreases with $k_{0}$ and takes its maximum value
for $k_{0}\cong\Lambda/3$, implying that this kind of interaction produce more decoherence for
small values of $k_0$ and, from the above discussion, for small values of $\Lambda$.
\begin{figure}
\begin{center}
\includegraphics[width=8cm]{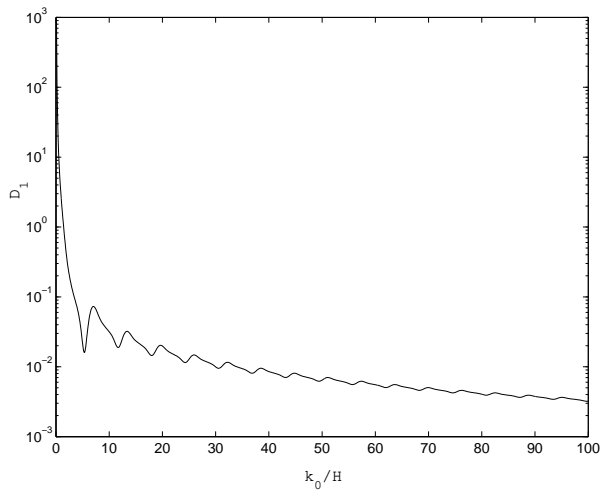}\\
\end{center}
\caption{ Coefficient ${D}_{1}$ (in logarithmic scale) as a function of  $k_{0}$ for
$H\eta=-0.5$.}\label{d1kG}
\end{figure} For practical purposes, we consider
the following simple approximation to $D_1$:
\begin{equation}\label{approxd1}
D_1^{\rm approx}(k_0 ,\eta,\Lambda)=-\frac{1}{100}\frac{(1 +
H\eta)}{H^4\eta^7{}k_{0}^{3}}\Theta(3{k}_{0}-\Lambda).
\end{equation}\begin{figure}
\begin{center}
\includegraphics[width=8cm]{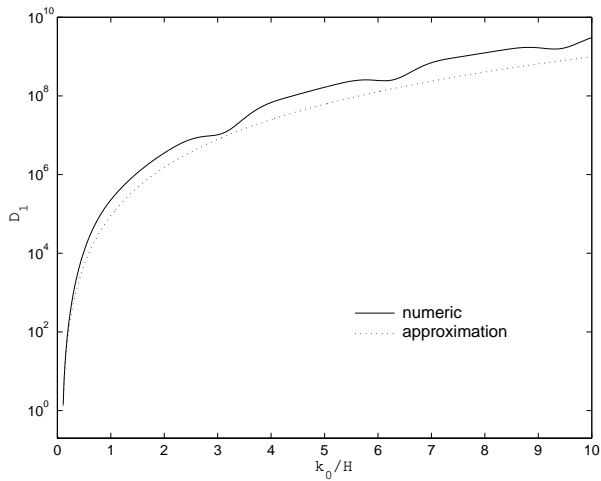}\\
\end{center}
\caption{Coefficient ${D}_{1}$ (in logarithmic scale) as a function of $k_{0}$ for a
fixed value of $|k_{0}\eta|=1/10$ ($k_{0}>H/10$), where we have
also plotted the approximation in
Eq.(\ref{approxd1}).}\label{d1scales}
\end{figure} As we can see from Fig. \ref{d1scales},
this approximation is less close to the exact coefficient for big
values of $k_0$. It is important to note from Eq.(\ref{deftdj}) that if the
approximation is a lower bound to the coefficient, then it will be  
useful to calculate an upper bound to $\eta_{d_1}$.

\begin{figure}
\begin{center}
\includegraphics[width=8cm]{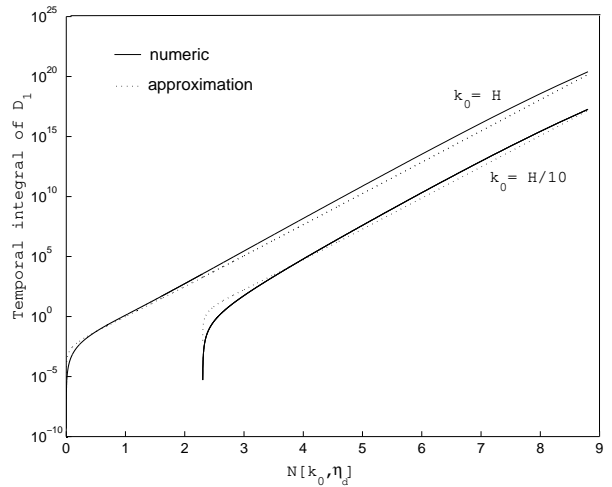}\\
\end{center}
\caption{Temporal integral of ${D}_{1}$ in units of $H^{-1}$ as a
function of $\mathcal{N}[k_{0},\eta_{d}]$ for two different values
of $k_{0}$. Solid curves are numerical integrations of the exact
coefficient, while short-dashed curves are from analytical
integration of Eq.(\ref{approxd1}). Note the logarithmic scale on the vertical axis.}\label{intd11yun10}
\end{figure}
According to the definition of $\eta_{d_{1}}$, given $k_{0}$,
$\Lambda$ and $\Gamma_{1}$, we can estimate this time by integrating $D_{1}$
 over the conformal time $\eta$ and plotting this temporal integral as a function of
$\mathcal{N}[k_{0},\eta_{d_{1}}]$. Fig. \ref{intd11yun10}
 shows such plots for two particular values of $k_0/H$, where
we have added the curves of the approximation (\ref{approxd1}). As
we have illustrated with these examples, the approximation to
$D_1$ is useful to estimate the order of magnitude of the
corresponding decoherence time $\eta_{d_1}$.

Let us now examine the behaviour of the coefficient $D_2$ which is
associated with the interaction term $\phi^{2}_{<}\phi_{>}^2$.
Since the interaction is now quadratic in $\phi_>$, there are no
restrictions on the values of $k_0$ such that $D_2\neq{}0$. Hence this coefficient
can affect the coherence of all modes in the system, therefore it is the most important
in our model.

\begin{figure}
\begin{center}
\includegraphics[width=8cm]{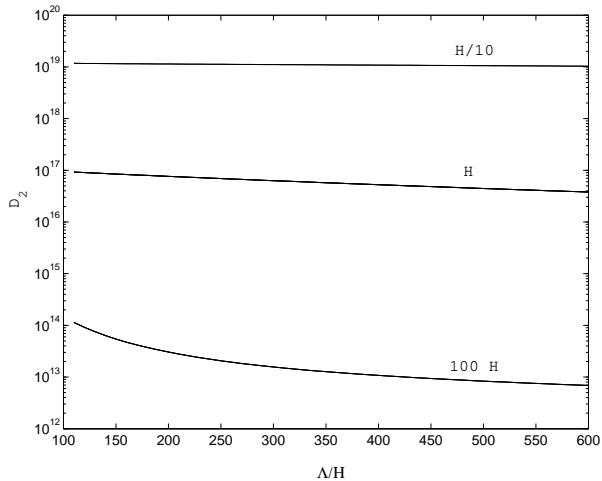}\\
\end{center}
\caption{ Coefficient ${D}_{2}$ (in logarithmic scale) as a function of
 $\Lambda/H$ for a fixed conformal time $\eta=-10^{-3}H^{-1}$ and three
 different values of $k_{0}$ ($H/10$, $H$ and $100 H$).}\label{d2vscut}
\end{figure}

The dependence of $D_2$ with $\Lambda$ is showed in Fig.
\ref{d2vscut} for three different values of $k_0$ and a fixed time
$\eta=-10^{-3}H^{-1}$. As we can see in this figure, for modes
with $k_{0}<<\Lambda$ the coefficient is weakly dependent of the
value of the critical wave vector $\Lambda$.

\begin{figure}
\begin{center}
\includegraphics[width=8cm]{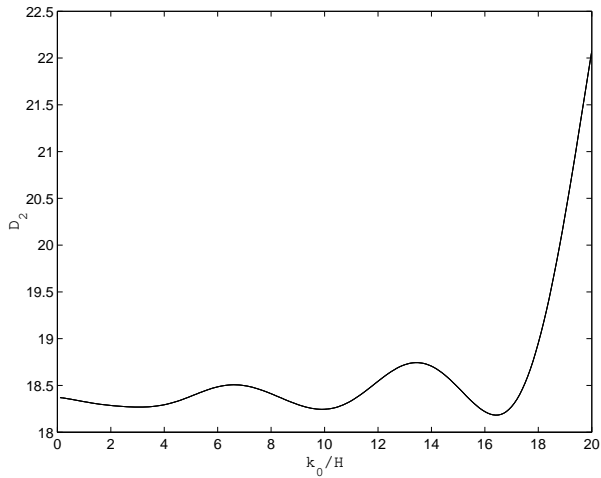}\\
\end{center}
\caption{Coefficient ${D}_{2}$ as a function of $k_{0}$ for $\Lambda=20{ }H$
and $H\eta=-1/2$.}\label{d2vskofuera}
\end{figure}

\begin{figure}
\begin{center}
\includegraphics[width=8cm]{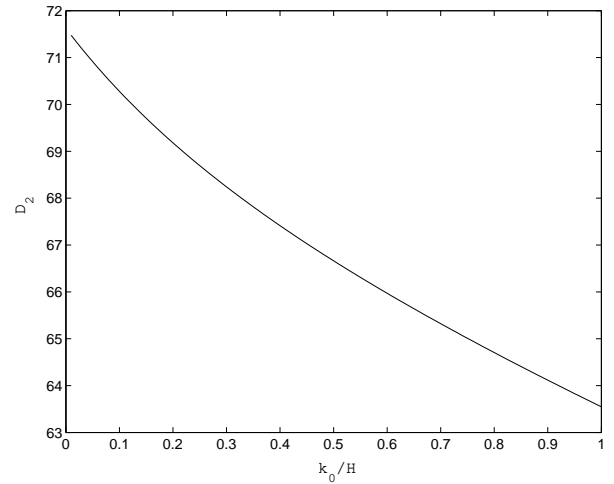}\\
\end{center}
\caption{ The same as Fig. \ref{d2vskofuera} but for modes $k_0$
outside the Hubble radius at
$\eta=-1/2{}H^{-1}$ and $\Lambda=2 H$.}\label{d2vskodentro}
\end{figure}

In Figs. \ref{d2vskofuera} and \ref{d2vskodentro} we have plotted
the coefficient $D_2$ as a function of $k_{0}/H$ for  fixed values
of $\eta$ and $\Lambda$. Fig. \ref{d2vskofuera} shows that
modes within the Hubble radius ($|k_0\eta|>1$) the diffusion
coefficient is an oscillatory function and it has a maximum
when $k_0 \sim \Lambda$. The same behaviour was noted for
conformally coupled fields \cite{lombmazz}. Physically, this fact
can be interpreted in terms of particle creation in the
environment due to its interaction with the system.
For these modes, a very simple but good approximation to $D_2$ is
given by
\begin{equation}\label{D2local}
D_2^{\ell}(\eta)=\frac{27}{2\pi}\frac{1}{(H\eta)^4},
\end{equation}
where the upper-script $\ell$ stands for ``local'', since it
corresponds to approximate the whole diffusion coefficient by the
term in Eq.(\ref{D2}) containing the Dirac delta function (see details in
Appendix \ref{AppendixA}). On the other hand, Fig. \ref{d2vskodentro} shows the
dependence of $D_2$ with $k_0$ for modes  outside the Hubble
radius at $\eta$. For these modes we note that the diffusive
effects are more important for the smallest values of $k_0$, which
are most sensitive at the expansion of the universe.
Note that the classical equation (\ref{newmodes}) for each mode
of the free field $\psi$ (defined as $\psi=a(\eta)\phi$) describes
an stable (unstable) oscillator if the mode $k_0$ satisfies
$|k_{0}\eta|>>1$ ($|k_{0}\eta|<<1$). Therefore, it was
expectable that the diffusion coefficients mirrors this fact.

For modes far outside ($|k_{0}\eta|<<1$) the Hubble radius we have
found an asymptotic approximation of $D_2$, useful if
$\Lambda\sim{}k_0$, which can be written as
\begin{equation}\label{D2aprox}
{D}^{a}_{2}(\eta,\Lambda)=-\frac{1}{\pi^{2}H^{4}\Lambda^{3}\eta^{7}}.
\end{equation}
\begin{figure}
\begin{center}
\includegraphics[width=8cm]{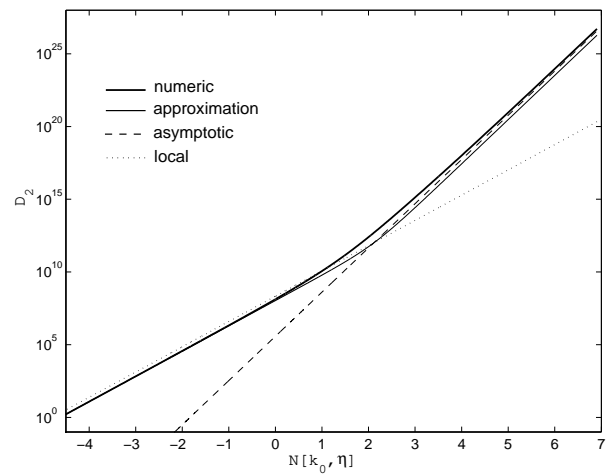}\\
\end{center}
\caption{Coefficient ${D}_{2}$ (in logarithmic scale) as a function of
$\mathcal{N}[k_{0},\eta_{d}]$ for $\Lambda=300{ }H$ and $k_{0}=100
H$. Heavy solid curve is numerically calculated from Eq.(\ref{D2}) 
(see details in Appendix \ref{AppendixA}). Light solid
curve corresponds to the approximation obtained by averaging the
local (dotted curve) and the asymptotic (short-dashed curve)
approximations.}\label{d2cienaprox}
\end{figure}
\begin{figure}
\begin{center}
\includegraphics[width=8cm]{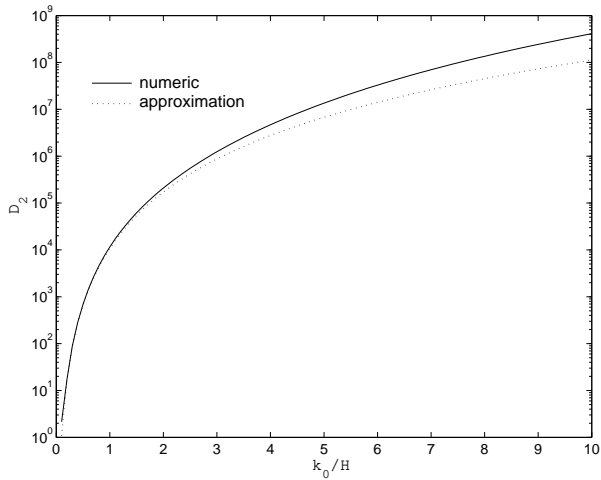}\\
\end{center}
\caption{ ${D}_{2}$ and $D_2^{approx}$ as functions of $k_{0}/H$
for a fixed value of $|k_{0}\eta|=1/10$
($k_{0}>H/10$) and $\Lambda=100{ }H$. The vertical axis is on a logarithmic scale.  }\label{d2vskoketafijo}
\end{figure}

\begin{figure}
\begin{center}
\includegraphics[width=8cm]{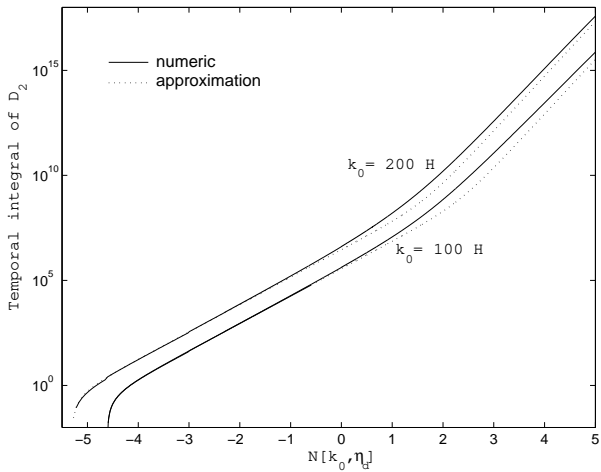}\\
\end{center}
\caption{Temporal integrals of ${D}_{2}$ and $D_2^{approx}$ (on a logarithmic scale and in units of $H^{-1}$) as
functions of $\mathcal{N}[k_{0},\eta_{d}]$, for $\Lambda=300{}H$
and two different values of $k_{0}$ ($100 H$ and $200
H$).}\label{intd2ciendos}
\end{figure}
\begin{figure}
\begin{center}
\includegraphics[width=8cm]{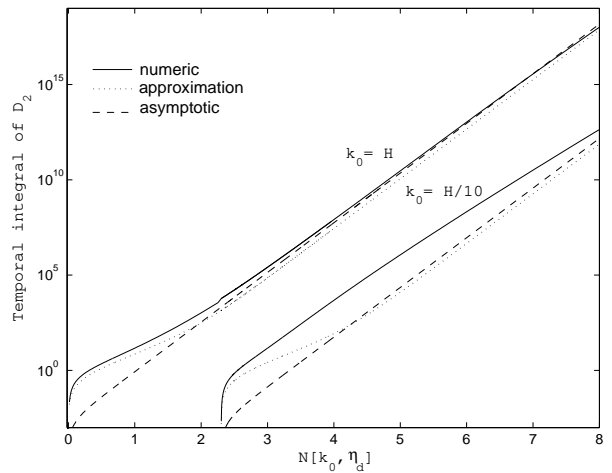}\\
\end{center}
\caption{The same as Fig. \ref{intd2ciendos} but for smaller
values of $k_{0}$ ($k_{0}=H$ and $k_{0}=H/10$) and
$\Lambda=2{}H$.}\label{intd2unounosodiez}
\end{figure}

Fig. \ref{d2cienaprox} shows a graph of the coefficient $D_2$ as a
function of $\mathcal{N}[k_{0},\eta_{d_{2}}]$ for a particular
value of $k_0$ and $\Lambda$, and the curves of the two
approximations above. From this figure, we can distinguish two
different regimes: one is well described by the local
approximation ($|k_{0}\eta|>1$) and the other by the asymptotic
one ($|k_{0}\eta|<<1$). This behaviour of $D_2$ indicates that the
decoherence process becomes faster once the mode crosses the
Hubble radius. In addition, we can see that a better approximation
to $D_2$ is obtained by averaging the local and the asymptotic
ones, that is
\begin{equation}\label{D2approx}
 D_2^{approx}=\frac{D_2^a+D_2^\ell}{2},
\end{equation}which is also showed in the same figure.

The dependence of $D_2$ with $k_0$ for a fixed value of
$\mathcal{N}[k_{0},\eta]$ is shown in Fig. \ref{d2vskoketafijo},
in which we see that the approximation is less close to the
numerical curve for big values of $k_0$, but it bounds $D_2$ from
below.

In Figs. \ref{intd2ciendos} and \ref{intd2unounosodiez} we have
plotted the temporal integral of the coefficient $D_2$ as a
function of $\mathcal{N}[k_{0},\eta_{d_{2}}]$ for different values
of $k_0$ and $\Lambda$, where we have also plotted the curves
obtained using the approximate expression given in Eq.(\ref{D2approx}). In view
of our analysis, it is reasonable to use that approximation to
estimate the decoherence time-scale $\eta_{d_2}$, at least for
values of $\Lambda$ smaller than $10^3{}k_0$.
\subsection{The decoherence time}

So far in this Section we have analyzed the behaviour of the
diffusion coefficients, and we have shown the approximations
considered are useful to estimate the decoherence time associated
with each diffusion coefficient (or interaction
term). Let us now work at the level of the order of magnitude and
apply these results to quantify the decoherence time $\eta_d$.

As we have already mentioned, $\eta_{d_1}$ and
$\eta_{d_2}$ are upper bounds to $\eta_d$. The time-scale given by
$\eta_d(k_0)$ sets the time after which we are able to distinguish
between two different amplitudes of the Fourier mode $k_0$ within
the volume $V$. Thus, the maximum value of $\eta_d(k_0)$ with
$k_0<\Lambda$ corresponds to an upper bound to the decoherence
time for the system-field (since, in principle, $\phi_<$ contains
all these modes with amplitudes different from zero).

In order to quantify decoherence times $\eta_{d_{1,2}}$ we have to
fix the values of $\Gamma_{1,2}$ (i.e., we have to assume values
to $\lambda$, $V$, $\phi_{<f}^+$, and $\phi_{<f}^-$). For this, as
a first approximation and since we are considering a fixed de
Sitter background, we will assume that the slow-roll conditions
are satisfied at least up to times of the order of the decoherence
time-scale. We will choose typical values for the parameters of
the model ($\lambda, V$) and for the elements of the reduced
density matrix ($\phi_{<f}^{+}$,$\phi_{<f}^{-}$).

The slow-roll conditions are usually written as
\cite{Peacock,langlois}:
\begin{equation}\label{SRparam}
\varepsilon_{U}=\frac{m^{2}_{\rm pl}}{16\pi}\left(\frac{U'}{U}\right)^{2}<<1\;,\;
\eta_{U}=\frac{m^{2}_{\rm pl}}{8\pi}\left(\frac{U''}{U}\right)<<1,
\end{equation}
where $m_{\rm pl}\equiv{}G^{-1/2}$ is the Plank mass,
$U=\lambda\phi_0^4$,  $U'=dU/d\phi_0$, and $\phi_0$ is the classical
inflaton field. If we also assume that
$\phi_0$ is homogeneous on physical scales
$\lambda_{\rm ph}>>({\sqrt{\lambda}\phi_{0}})^{-1}$, these conditions
imply that the classical configuration field $\phi_0$ and the
Hubble rate satisfy
\begin{equation}\label{eqslowroll}  H^{2}\simeq\frac{8\pi{}U}{3{}m_{\rm pl}^{2}}~\;;\;
~\frac{d\phi_0}{dt}\simeq-\frac{U'}{3{}H}.
\end{equation}

Defining the end of the inflationary period setting $\epsilon_U
\sim 1$, one can set
$\phi_0(N_\eta)\cong\sqrt{(N_\eta+1)/{\pi}}\;m_{\rm pl}$, where
$N_\eta = \ln{a(\eta_f)/a(\eta)}$; typically the e-fold number
$N_{\eta_i}\equiv{}N\sim 60$ \cite{langlois}.  Thus, we assume the
mean value of the system-field at time of decoherence is
$\phi_0(N_{\eta_d})$ ($\Sigma_{\phi_{<f}}\equiv(\phi_{<f}^+ +
\phi_{<f}^-)/2\simeq\phi_0(N_{\eta_d})$). Using this mean value
and Eq.(\ref{eqslowroll}) we can write
$H^{2}\simeq{}8\lambda{}N_{\eta_d}^{2}m_{\rm pl}^{2}/(3\pi)$. In
order to choose a typical order of magnitude for the fluctuations
of the system-field, let us use the fact that the amplitude of the
so-called primordial density perturbations $\delta$ can be
inferred to be of order $10^{-5}$, and that $\delta\sim
H/\dot{\phi_0}\delta\phi$, where $\delta\phi$ is the amplitude of
the inflaton field fluctuations \cite{Peacock,langlois}. Thus,
with the use of these constraints and Eq.(\ref{eqslowroll}) we set
$\Delta_{\phi_{<f}}\equiv(\phi_{<f}^+ - \phi_{<f}^-)\sim
10^{-5}\phi_0(N_{\eta_d})/N_{\eta_d}$. Since $V$ is the spatial
volume inside which there are no coherent superpositions of
macroscopically distinguishable states for the system-field, it is
reasonable to choose $V= v H^{-3}$, with $v\sim{}1$ (which
corresponds, if $N\sim{}60$, to the comoving volume $V \sim
(a_0{}H_0)^{-3}$, where $H_0$ and $a_0$ are the present values of
the Hubble rate and the scalar factor, respectively) .

From previous considerations and according to the condition
(\ref{deftdj}), the times $\eta_{d_{1,2}}$ can be estimated as the
solution to
\begin{subequations}\label{est}
 \begin{align}\label{est1}
 \int_{\eta_{i}}^{\eta_{d_{1}}}H{}d\eta{D}_{1}(\eta)&\sim\frac{H}{\Gamma_1}\lesssim{}2\times{}1
   0^{15}\frac{(\lambda{}10^{5})}{v}\left(\frac{N}{60}\right)^{5}\\\label{est2}
 \int_{\eta_{i}}^{\eta_{d_{2}}}H{}d\eta{D}_{2}(\eta)&\sim\frac{H}{\Gamma_2}\lesssim{}9\times{}10^
  {17}\frac{1}{v}\left(\frac{N}{60}\right)^{4},\;\;\;
\end{align}
\end{subequations}
where we have used the fact that $N_{\eta_d}\lesssim{}N$. The last
term in (\ref{est1}) follows after taking
$(\phi_{<f}^{+3}-\phi_{<f}^{-3})\sim{}\Delta_{\phi_{<f}}\Sigma^{2}_{\phi_{<f}}$.
For example, from Fig. \ref{intd2unounosodiez} we can see that the
temporal integral of $D_2$ takes a value of order $10^{17}$ for
$\mathcal{N}[k_0=H,\eta_{d_{2}}]$ between $7$ and $8$. As $D_2$ is
weakly dependent with $\Lambda$ for $k_0<H$, we can say that it is
valid for $\Lambda\sim{}H$. Thus, since $D_2$ decrease with $k_0$
for $|k_0\eta|<1$, we can conclude that the decoherence time for
those modes with $k_0<H$ are smaller.

Substituting the approximation given in Eq.(\ref{approxd1}) into
the left-hand side of (\ref{est1}) and assuming that
$|H\eta_{d_{1}}|<<1$, we get
\begin{eqnarray}\label{estimaciontd1}
t_{d_{1}}&\sim&\frac{1}{6H}\ln\left(600\alpha^3\frac{H}{\Gamma_1}\right)\\
\nonumber
&\lesssim&\frac{7}{H}+\frac{1}{6{}H}\ln\left(\lambda{}10^{5}\right)+
\frac{1}{6{}H}\ln\left(\frac{\alpha^{3}}{v}\left(\frac{N}{60}\right)^{5}\right),
\end{eqnarray}
where $t_{d_{1}}=H^{-1}\ln a(\eta_{d_1})$, and we have defined $k_0=\alpha{}H$.

With the use of the approximation in Eq.(\ref{D2approx}) we obtain
a quadratic equation in
$x\equiv{}a^3(\eta_{d_2})=\exp(3{}H{}t_{d_2})$, which is simple to
solve for $t_{d_2}$. The result is
\begin{eqnarray}\label{cuadratica}\nonumber
x&\sim&\frac{27}{4}\pi{}\sigma^3{}\left(\sqrt{1+\frac{8}{27\sigma^3}\left[\frac{8{}H}{9{}\Gamma_2}+
\frac{1}{\pi}+\frac{2}{27\pi^2\sigma^3}\right]}-1\right)\\
&<&\frac{27}{4}\pi{}\sigma^3{}\left(\sqrt{1+\frac{8}{27\sigma^3}\left[\frac{8{}H}{9{}\Gamma_2}+
\frac{1}{\pi}+\frac{2}{27\pi^2\sigma^3}\right]}\right),
\end{eqnarray}
where $\sigma\equiv{}\Lambda/H$. For the sake of simplicity let us
set $\sigma\sim{}1$. Since ${H}/{\Gamma_2}>>1$, it yields
\begin{eqnarray}\label{estimaciontd1}\nonumber
t_{d_{2}}&\lesssim&\frac{1}{6H}\ln\left(12\pi^2\sigma^3\frac{H}{\Gamma_2}\right)\\
&\lesssim&\frac{7.7}{H}+\frac{1}{6{}H}\ln\left(\frac{\sigma^3}{v}\left(\frac{N}{60}\right)^4\right).
\end{eqnarray}Assuming  $N = H t_{\rm end} \geq 60$ as an estimative scale to
the end of inflationary period and values of $\lambda \leq
10^{-5}$, we obtain
\begin{equation}
\frac{t_{d_1}}{t_{\rm end}}\lesssim
\frac{7}{60}+\frac{1}{120}\ln\left(\frac{\alpha}{v^{1/3}}\right),
\end{equation}
which makes sense only if $k_0<\Lambda<3{}k_0$
($\alpha<\sigma<3{}\alpha$), and
\begin{equation}
\frac{t_{d_2}}{t_{\rm end}}\lesssim 
\frac{2}{15}+\frac{1}{120}\ln\left(\frac{\sigma}{v^{1/3}}\right).
\end{equation}

From scales $t_{d_1}$ and $t_{d_2}$ we conclude that if one set
$\Lambda \lesssim H$, the decoherence time-scale for the
system-field is shorter than the minimal duration of inflation for
all the wave-vectors within the system sector.

\section{Effective dynamical evolution of the system-field in a fixed de Sitter background}

In this Section we concern ourselves with the time evolution
of the system-field. After reviewing the phenomenological way to
describe the stochastic dynamical evolution of the system, we present the renormalized 
``semiclassical-Langevin'' equation. This equation can be used
to describe the dynamical evolution of the classical
configurations of the system-field and hence is useful once
all the modes in the system have lost coherence.  As a first
step to understand the generation of classical inhomogeneities
from quantum fluctuations, we then consider a simple
situation in which we analyze the influence of the environment on
the power spectrum for modes inside the system sector.

\subsection{Effective dynamical evolution, noise and expectation values}

The real part of the influence action contains divergent terms and should be renormalized. The 
imaginary part is finite and is associated with the decoherence process.  It is well known that the 
terms of the imaginary part that come from a given interaction term in
the original action can be viewed as arising from a noise source \cite{lomplb, GMuler}. In our case 
there are two such sources $\xi_2$ and $\xi_3$, which are associated with the interaction terms 
$\phi^2_{<}\phi^2_{>}$ and $\phi^3_{<}\phi_{>}$ respectively.
That is, the imaginary part of the influence action can be rewritten as
\begin{equation}
Im{\delta{}A}=-\ln\left(F[\Delta_2]F[\Delta_3]\right),
\end{equation}
where $F[\Delta_n]$ ($n=2,3$) is the characteristic functional of the noise $\xi_n$,
which is related with Gaussian functional probability distribution $P[\xi_n]$ as
\begin{eqnarray}\nonumber
F[\Delta_{n}]&=&\int{D}\xi_{n}\;P[\xi_{n}]\exp\left\{-i\int{}d^{4}x\Delta_{n}(x)\xi_{n}(x)\right\},
\\\label{PP}\nonumber
P[\xi_{n}]&=&{}N_{n}\exp\Big{\{}-\frac{1}{2}\int{d}^{4}x_{1}\int{}d^{4}x_{2}\;\xi_{n}(x_{1})\\
&\times&\nu^{-1}_{n}(x_{1},x_{2})\xi_{n}(x_{2})\Big{\}},
\end{eqnarray}where  $N_n$ is a normalization factor and $\nu_n^{-1}$ is the functional 
inverse of the noise kernel $\nu_n$:
\begin{subequations}\begin{align}\label{nu}
\nu_{2}(x_{1},x_{2})&=-288\lambda^{2}a^{4}(\eta_{1})a^{4}(\eta_{2})ReG^{\Lambda 2}_{++}(x_{1},x_{2}),\\
\nu_{3}(x_{1},x_{2})&=64\lambda^{2}a^{4}(\eta_{1})a^{4}(\eta_{2})ImG^{\Lambda}_{++}(x_{1},x_{2}).
\end{align}\end{subequations}
The Gaussian noise field $\xi_n(x)$ is completely characterized by
\begin{subequations}\begin{align}
\langle\xi_{n}(x)\rangle_{P}&=0,\\
\langle\xi_{n}(x_{1})\xi_{n}(x_{2})\rangle_{P}&=\nu_{n}(x_{1},x_{2}),
\end{align}\end{subequations} 
where  with $\langle\rangle_{P}$  we are denoting average over all realizations of $\xi_n(x)$.

The functional variation
\begin{equation}\label{fvar}
\frac{\delta{S_{\rm eff}}}{\delta\phi^{+}_{<}}\Big{|}_{\phi^{+}_{<}=\phi^{-}_{<}}=0,
\end{equation}
yields the ``semiclassical-Langevin'' equation for the
system-field, which is only valid once the system has become
classical.

With the identifications above, the reduced density matrix can be rewritten as \cite{BLHU}:
\begin{eqnarray}\label{evolucionMDRepromstoc}
\rho_{\rm r}[\phi^{+}_{<f}|\phi^{-}_{<f};\eta]&=&\int{}{D}\xi_{2}P[\xi_{2}]\int{{D}}\xi_{3}P[\xi_{3}]
\\\nonumber
&\times&\rho_{\rm r}[\phi^{+}_{<f}|\phi^{-}_{<f},\xi_{2},\xi_{3};\eta],
\end{eqnarray} with
\begin{subequations}\begin{align}\label{evolucionMDRestoch}\nonumber
&\rho_{\rm r}[\phi^{+}_{<f}|\phi^{-}_{<f},\xi_{2},\xi_{3};\eta]\equiv\int{d}\phi^{+}_{<i}\int{d}
\phi^{-}_{<i}\int_{\phi^{+}_{<i}}^{\phi^{+}_{<f}}{D}\phi^{+}_{<}\\\;\;\nonumber
&\times\;\int_{\phi^{-}_{<i}}^{\phi^{-}_{<f}}{D}\phi^{-}_{<}\;
\rho_{\rm r}[\phi^{+}_{<i}|\phi^{-}_{<i};\eta_{i}]\exp\left\{i{}S_{\rm eff}[\phi^{+}_{<},
\phi^{-}_{<},\xi_{2},\xi_{3}]\right\},
\end{align}\end{subequations} where the effective action $S_{\rm eff}$ is given by
\begin{eqnarray}\label{Seff}
S_{\rm eff}[\phi^{+}_{<},\phi^{-}_{<},\xi_{2},\xi_{3}]&=&Re\{A[\phi^{+}_{<},\phi^{-}_{<}]\}\
\\\nonumber
&-&\int{d^{4}x}[\Delta_{2}(x)\xi_{2}(x)+\Delta_{3}(x)\xi_{3}(x)].
\end{eqnarray} Here $A$ is the CGEA specified in Eq.(\ref{CTPEA}). Thus, the full expectation value of any operator $\hat{Q}[\phi_<]$ can be written as
\begin{eqnarray}\label{valmed}\nonumber
\langle\hat{Q}[\phi_{<}]\rangle&=&\int{{D}}\xi_{2}P[\xi_{2}]\int{{D}}\xi_{3}P[\xi_{3}]\int{d}\phi_{<}
\\\nonumber&\times&\;\rho_{\rm r}[\phi_{<}|\phi_{<},\xi_{2},\xi_{3};\eta]\;{Q}[\phi_{<}]\\
&\equiv& \langle\langle\hat{Q}[\phi_{<}]\rangle_q\rangle_P,
\end{eqnarray} where $\langle\rangle_{q}$ is an usual quantum average for a system-field  
subjected to external stochastic forces.

\subsection{Renormalized equation of motion for the system}

Taking the functional variation as in Eq.(\ref{fvar}) we obtain

\begin{eqnarray} \label{ecSClasDS}\nonumber
&&\phi_{<}''(\eta,\vec{x})-\triangle\phi_{<}(\eta,\vec{x})+2\;\mathcal{H}\phi'_{<}(\eta,\vec{x})\\\nonumber
&&+\;4\;\lambda{}a^{2}(\eta)\phi_{<}(\eta,\vec{x})[\phi^{2}_{<}(\eta,\vec{x})-3iG^{\Lambda}_{++}
(\eta,\eta,\vec{0})]\\
\nonumber
&&-\;96\;\lambda^{2}{a}^{2}(\eta)\int_{\eta_{i}}^{\eta}d\eta'{}a^{4}(\eta'){
}\phi^{2}_{<}(\eta,\vec{x})\\\nonumber
&&\times\int{}d^{3}y\phi^{3}_{<}(\eta',\vec{y})ReG^{\Lambda}_{++}(\eta,\eta',\vec{x}-\vec{y})\\
\nonumber
&&-\;288\;\lambda^{2}
{a}^{2}(\eta)\int_{\eta_{i}}^{\eta}d\eta'{}a^{4}(\eta'){
}\phi_{<}(\eta,\vec{x})\\\nonumber
&&\times\int{}d^{3}y
\phi^{2}_{<}(\eta',\vec{y})ImG^{\Lambda 2}_{++}(\eta,\eta',\vec{x}-\vec{y})\\
&&=-\xi_{2}(\eta,\vec{x})\frac{\phi_{<}(\eta,\vec{x})}{a^{2}(\eta)}-\frac{3}{2}\xi_{3}(\eta,\vec{x})
\frac{\phi^{2}_{<}(\eta,\vec{x})}{a^{2}(\eta)}.
\end{eqnarray}
This equation contains divergences. In
order to renormalize it we use the method of adiabatic substraction
with dimensional regularization, which works at the level of the
field equation and is particularly useful for solving the equation
numerically \cite{RenoPazMazziCarmen}. To simplify the task we
assume the system-field to be homogeneous enough so that the
spatial derivatives and the term in
Eq.(\ref{ecSClasDS}) coming from the interaction
$\phi_{>}\phi^{3}_{<}$ are negligible. Note that for an homogeneous system-field there is no contribution from this interaction term up to one loop order, due to orthogonality of the Fourier modes (see, for instance, Eq.(\ref{inffimag})). Details of the
renormalization procedure are relegated to Appendix
\ref{AppendixB}. The renormalized equation is
\begin{eqnarray}\label{ERen}
&&\phi_{<}''(\eta)+[\Delta{}M^{2}(\eta)+\Delta{}\Sigma(\eta)R]\;a^{2}(\eta)\;\phi_{<}(\eta)\\\nonumber
&&+\;2\;\mathcal{H}\;\phi'_{<}(\eta)+4\;[\lambda+\Delta\tilde{\lambda}(\eta)]\;a^{2}(\eta)\;
\phi^{3}_{<}(\eta)\\\nonumber
&&+\;\frac{36\lambda^{2}}{\pi^{2}}\phi_{<}(\eta)a^{2}(\eta)\left\{\int_{\eta_{i}}^{\eta}d\eta'\;
\phi_{<}^{2}(\eta')\mathcal{J}(\eta,\eta')\right.\\\nonumber
&&+\left.\int_{\eta_{i}}^{\eta}d\eta'\frac{\phi_{<}(\eta'){\phi'}_{<}(\eta')}{H{}a(\eta)}
\mathcal{R}(\eta,\eta')\right\}=-\tilde{\xi}_{2}(\eta)\frac{\phi_{<}(\eta)}{a^{2}(\eta)}.
\end{eqnarray}where $R=12 H^2$; we have redefined the noise source as
\begin{equation}\label{ruru}
 \tilde{\xi}_{2}(\eta)=\frac{1}{V}\int_{V}d^{3}x\;\xi_{2}(\eta,\vec{x}),
\end{equation} whose correlation function is given by\begin{eqnarray}\nonumber
&&\langle\tilde{\xi}_{2}(\eta_{1})\tilde{\xi}_{2}(\eta_{2})\rangle_{P}= a^{2}(\eta_{1})a^{2}
(\eta_{2})\frac{36\lambda^{2}}{\pi^{2}V}\left\{\frac{\pi}{2}\;\delta[(\eta_{1}-\eta_{2})] \right.\\
\nonumber
&&+\frac{\cos[2\Lambda(\eta_{1}-\eta_{2})]}{\Lambda{}}\left[\frac{2}{\eta_{1}\eta_{2}}+
\frac{(\eta_{1}-\eta_{2})^{2}}{3\;\eta_{1}^{2}\eta_{2}^{2}}+\frac{1}{3\;
\Lambda^{2}\eta_{1}^{2}\eta_{2}^{2}}\right]\\
\nonumber
&&-(\pi-2\;\mathcal{S}i[2\Lambda|\eta_{1}-\eta_{2}|])\left[\frac{|\eta_{1}-\eta_{2}|^{3}}
{3\;\eta_{1}^{2}\eta_{2}^{2}}+\frac{|\eta_{1}-\eta_{2}|}{\eta_{1}\eta_{2}}\right]\\\nonumber
&&\left.-\sin[2\Lambda(\eta_{1}-\eta_{2})]\left[\frac{1}{2(\eta_{1}-\eta_{2})}-
\frac{2(\eta_{1}-\eta_{2})}{3\;\Lambda^{2}\eta_{1}^{2}\eta_{2}^{2}}\right]\right\};
\end{eqnarray}and we have also defined the following functions:

\begin{subequations}\label{addterms}\begin{align}
\Delta{}\tilde{\lambda}(\eta)&=\Delta{}\lambda+\frac{9\lambda^{2}}{2\pi^{2}}\left[\frac{\gamma}{2}-
\ln\Big{|}\frac{a(\eta)\mu}{2\Lambda}\Big{|}\right],\\
\Delta{}\Sigma(\eta)&=\Delta{}\xi-\frac{\lambda}{4\pi^{2}}\left[\frac{\gamma}{2}-
\ln\Big{|}\frac{a(\eta)\mu}{2\Lambda}\Big{|}\right],\\
\Delta{}M^{2}(\eta)&=\Delta{}m^{2}-\frac{18\lambda^{2}}{\pi^{2}}\frac{a^{2}(\eta_{i})}{a^{2}(\eta)}
\phi_{<}^{2}(\eta_{i})\\\nonumber
&\times\mathcal{C}i[2\Lambda(\eta-\eta_{i})]-\frac{3\lambda\Lambda^{2}}{2\pi^{2}a^{2}(\eta)},\\
\mathcal{R}(\eta,\eta')&=\frac{\eta}{{\eta'}^{2}}\mathcal{C}i[2\Lambda(\eta-\eta')],\\
\mathcal{J}(\eta,\eta')&=\mathcal{J}_{1}(\eta,\eta')+\mathcal{J}_{2}(\eta,\eta')+
\mathcal{J}_{3}(\eta,\eta'),\\
\mathcal{J}_{1}(\eta,\eta')&=\mathcal{C}i[2\Lambda(\eta-\eta')]\left[\frac{2(\eta^{3}-
{\eta'}^{3})}{3{\eta'}^{4}}+\frac{\eta^{2}}{{\eta'}^{3}}\right],\\
\mathcal{J}_{2}(\eta,\eta')&=\frac{2(\eta-\eta')}{3{\eta'}^{4}\Lambda^{2}}
\left[\cos[2\Lambda(\eta-\eta')]\right.\\\nonumber
&\left.-\frac{\sin[2\Lambda(\eta-\eta')]}{2\Lambda(\eta-\eta')}\right],\\
\mathcal{J}_{3}(\eta,\eta')&=-\frac{\sin[2\Lambda(\eta-\eta')]}{\Lambda}
\left[\frac{(\eta-\eta')^{2}}{3{\eta'}^{4}}+\frac{2\eta}{{\eta'}^{3}}\right],
\end{align}
\end{subequations} where $\mathcal{C}i[x]$ is the cosine integral function \cite{Abramob}.
 Notice that these functions
are logarithmically divergent in the limit $\Lambda\rightarrow{}0$, which is the well-known 
infrared divergence \cite{RenoPazMazziCarmen}.

\begin{figure}
\begin{center}
\includegraphics[width=8cm]{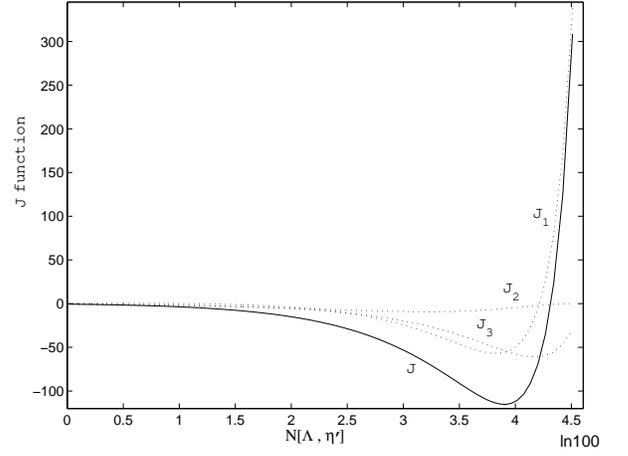}\\
\end{center}
\caption{$\mathcal{J}$ and $\mathcal{J}_{1,2,3}$ as functions of
$\mathcal{N}[\Lambda,\eta']\equiv-\ln|\Lambda\eta'|$ for
$\Lambda\eta=-10^{-2}$
($\mathcal{N}[\Lambda,\eta]=\ln{100}\simeq{}4.605$).}\label{Jotas}
\end{figure}

The functions above contain useful information. Particularly, they allows us to examine
 the   conditions under which the loop expansion breaks down. In order to estimate the
  time after which the one-loop terms become of the same order of magnitude as the classical
   ones we may compare their time-dependent parts. For example, from Eq.(\ref{addterms}) we
   see that the time-dependent part of $\Delta{}\tilde{\lambda}(\eta)$ ($\Delta{}\Sigma(\eta)$)
   is the order $\lambda$ (one) for $H{}t\sim1/\lambda$ and $\Delta{}M^{2}(\eta)$ is important
   only at the initial time ($\eta\sim\eta_i$). In the same way,  we can use the term
   $\lambda{}a^2\phi_<^3$ to compare it with the remaining ones.
In Fig. \ref{Jotas} we have plotted the $\mathcal{J}_{i}$
functions, where we can see that $\mathcal{J}_{1}$ dominates all
others. It allow us to make the following approximation:
\begin{eqnarray}\label{ApJ}\nonumber
\int_{\eta_{i}}^{\eta}d\eta'\;\phi_{<}^{2}(\eta')\mathcal{J}(\eta,\eta')&\simeq&
\phi_{<}^{2}(\eta)\int_{\eta_{i}}^{\eta}d\eta'\;\mathcal{J}_{1}(\eta,\eta')\\
&\sim&-\phi_{<}^{2}(\eta)\frac{1}{3}(\ln|\Lambda\eta|)^{2},
\end{eqnarray}
where the last term is a simple long-time asymptotic expression.
With the use of this approximation we obtain that the two terms in
question are of the same order when
 $|\ln|\Lambda\eta||\sim{}1/\sqrt{\lambda}$ and,
for $\Lambda\sim{}H$, $H{}t\sim{}1/\sqrt{\lambda}$.

\begin{figure}
\begin{center}
\includegraphics[width=8cm]{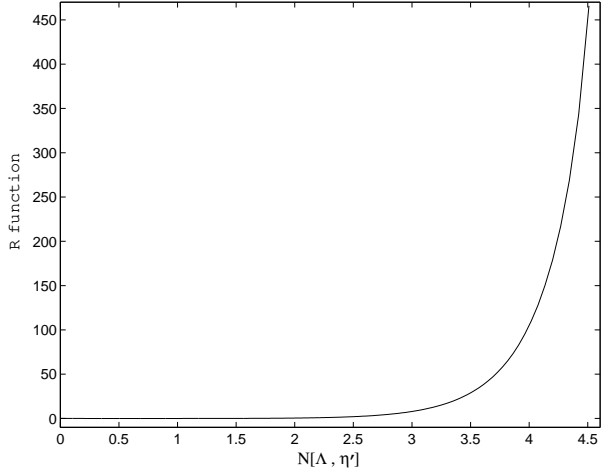}\\
\end{center}
\caption{$\mathcal{R}$ as a function of $\mathcal{N}[\Lambda,\eta']$
for $\Lambda\eta=-10^{-2}$
($\mathcal{N}[\Lambda,\eta]=\ln{100}\simeq{}4.605$).}\label{R}
\end{figure}

As it is shown in Fig. \ref{R}, the $\mathcal{R}$ function peaks at
$\eta\sim\eta'$ and hence we can
approximate\begin{eqnarray}\label{aplocR}
\int_{\eta_{i}}^{\eta}d\eta'\frac{\phi_{<}(\eta'){\phi'}_{<}(\eta')}{H{}a(\eta)}\mathcal{R}&
\simeq&\frac{\phi_{<}(\eta){\phi'}_{<}(\eta)}{H{}a(\eta)}
\int_{\eta_{i}}^{\eta}d\eta'\mathcal{R}\\
\nonumber&\sim&-\frac{\phi_{<}(\eta){\phi'}_{<}(\eta)}{H{}a(\eta)}\nonumber \\
&\times & (\gamma+\ln|2\Lambda\eta|),
\end{eqnarray}where the last expression corresponds to a long-time approximation. As for the 
$\mathcal{J}$ term, using
 the long-time approximation we get $\ln|\Lambda\eta|\sim\phi_{<}H/(\lambda{
}\dot{\phi}_{<})$. If in addition we assume that the slow-roll
conditions are satisfied, this time can be estimated as
$|\ln|\Lambda\eta||\sim{1}/(\lambda{}\varepsilon_{U})$ and,
for $\Lambda\sim{}H$,
$H{}t\sim{1}/(\lambda{}\varepsilon_{U})$,
where $\varepsilon_{U}$ is the slow-roll parameter given in
Eq.(\ref{SRparam}).

From previous discussion, we can see that the additional terms are
only important at times longer than the typical time-scale 
associated with the decoherence process.

\subsection{Generation of inhomogeneities: role of the noise}

Having reached this point, we ask ourself about the primordial
inhomogeneities. Certainly, the model we are considering is too
simplified to obtain any property of the power spectrum of such
inhomogeneities. Nevertheless, we can discuss some relevant
aspects about the way of connecting the initial quantum
fluctuations with those which eventually become classical and
statistical. With this purpose, let us compute the power spectrum
for some of the Fourier modes in the system working in an
approximate way and making certain assumptions, but taking into
account the decoherence process.

According to the results of the previous Section, the decoherence
process occurs at different rates for each mode in the system.
Provided that decoherence is  more effective for modes outside the
Hubble radius, we consider the following situation. We suppose
that the ``relevant'' modes are in the Bunch-Davies state (i.e.,
in equilibrium with the environment) at the initial time $\eta_i$.
Consequently, the matrix elements of the initial density operator
for each of these modes are Gaussian functions in a mode-amplitude
basis. As a first approximation we neglect the non-linearities
which might affect the Gaussian form of those matrix elements.

We are now interested only in computing the power spectrum of
those ``relevant'' modes up to $\hbar$ and $\lambda^2$ order. To
carry this out, we split the system-field as
$\phi_{<}=\phi_{0}(\eta)+\delta\phi_{<}$, where we identify
$\phi_{0}(\eta)$ as a classical background field which satisfies
slow-roll conditions. The power spectrum of the quantum field fluctuations
$\delta\phi_<$ may be expressed as
$P_{\phi}(k)=2\pi^{2}k^{-3}\Delta^{2}_{\phi}(k)$, with
$\Delta^{2}_{\phi}(k)$  defined by
\begin{equation}\label{pkcampo}
\langle{\delta\phi_<}(\vec{x}){\delta\phi_<}(\vec{x}+\vec{r})\rangle=\int{}d^{3}k\;
\frac{\Delta_{\phi}^{2}(k)}{4\pi{}k^{3}}\exp(-i\vec{k}\cdot\vec{r}),
\end{equation} where $\langle\rangle$ is the full average  in
Eq.(\ref{valmed}).

Consistently with the assumption that the ``relevant'' matrix
elements remain Gaussian, we expand the semiclassical equation
(\ref{ecSClasDS})  up to linear order in the mode amplitude of
interest $\delta\phi_<(\vec{k})$. Through this procedure we obtain
\begin{subequations}
    \begin{align}\label{ecuacionclasica}
&\phi_{0}''(\eta)+2\mathcal{H}\phi_{0}'(\eta)+4\lambda{}a^{2}\phi_{0}^{3}(\eta)=0,\\\label{ecuacionmodos}
&\delta\phi_{<}''(\vec{k},\eta)+[k^{2}+12\lambda{}a^{2}\phi_{0}^{2}(\eta)]\delta\phi_{<}(\vec{k},\eta)
\nonumber\\
&+2\mathcal{H}\delta\phi_{<}'(\vec{k},\eta)=-\frac{\xi_{2}(\vec{k},\eta)}{a^{2}}\phi_{0}(\eta),
 \end{align}
\end{subequations}where we have discarded the terms which do
not contribute to the power spectrum up to $\hbar$ order. The term
with  the $\xi_3$  noise  source gives a zero contribution due to
our approximations and the orthogonality of the Fourier modes. It
is important to note the presence of the $\xi_2$  noise  source,
which ensure the decoherence process occurs as the matrix elements
are evolved in time.

In order to obtain the power spectrum let us  split
$\delta\phi_{<}(\vec{k},\eta)=\delta\phi_{<}^{\xi}(\vec{k},\eta)+\delta\phi_{<}^{q}(\vec{k},\eta)$,
with
$\delta\phi_{<}^{\xi}(\vec{k},\eta)\equiv\langle\delta\phi_{<}(\vec{k},\eta)\rangle_{q}$,
where $\langle\rangle_{q}$ is the quantum average defined in
Eq.(\ref{valmed}). Because of the assumption of linearity, this
quantum average satisfies the semiclassical equation
(\ref{ecuacionmodos}), whose solution can be written as the sum of
the homogeneous solution $\delta\phi_{<}^{h}(k,\eta)$ and a
particular solution $\delta\phi_{<}^{p}(\vec{k},\eta)$. The former is given by
\begin{equation}\label{homog}
\delta\phi_{<}^{h}(k,\eta)=a^{-1}(\eta)[\alpha_{k}\sqrt{|\eta|}J_{\nu}+
\beta_{k}\sqrt{|\eta|}J_{-\nu}],\\
\end{equation}where $\alpha_{k}$ and $\beta_{k}$ are constants of integration, and 
$\nu=\sqrt{\frac{9}{4}-\epsilon}$, with
$\epsilon={6\lambda\phi_{0}^{2}}/{H^{2}}$.
Setting $\nu\simeq\frac{3}{2}$, a particular solution
is\begin{equation}\label{phipartic}
\delta\phi_{<}^{p}(\vec{k},\eta)=-\int_{\eta_{i}}^{\eta}d\eta'{}
g(k,\eta,\eta')\xi_{2}(\vec{k},\eta')\phi_{0}(\eta'),
\end{equation}where
\begin{eqnarray}\label{gpro}
g(k,\eta,\eta')&=&\frac{1}{a(\eta)a(\eta')}\left[\frac{\sin[k(\eta-\eta')]}{k}
\left(1+\frac{1}{k^{2}\eta\eta'}\right)\right.\nonumber\\
&-&\left.\frac{\cos[k(\eta-\eta')]}{k^{2}\eta\eta'}(\eta-\eta')\right].
\end{eqnarray}

With the use of the initial conditions
$\langle\delta\phi_{<}(\vec{k},\eta_{i})\rangle_{q}=\langle\dot{\delta\phi_{<}}(\vec{k},
\eta_{i})\rangle_{q}=0$,
we obtain that
$\langle\delta\phi_{<}^{\xi}(\vec{k},\eta)\rangle_{P}=0$, and thus
$\delta\phi_{<}^{\xi}(\vec{k},\eta)=\delta\phi_{<}^{p}(\vec{k},\eta)$.
Within these approximations, the result is analogous to that for
the linear quantum Brownian motion (QBM) \cite{qbm,Morika}.
Therefore, the quantity $\Delta_{\phi}^{2}(k)$
 can be written so that it receives two contributions:
\begin{equation}\label{res}
\Delta_{\phi}^{2}(k)=\Delta_{\phi^{q}}^{2}(k)+\Delta_{\phi^{\xi}}^{2}(k).
\end{equation} The first one comes
from the unitary evolution of the initial density matrix, i.e., it
is the usual quantum result for the case of the free field \cite{langlois}: 
$\Delta_{\phi^{q}}^{2}(k)=\left({H}/{2\pi}\right)^{2}(1+k^{2}\eta^{2})$.
The second one appears due to the $\xi_{2}$ noise source and can
 be computed through
\begin{eqnarray}\label{corr}\nonumber
\langle\delta\phi_{<}^{\xi}(\vec{k_{1}},\eta)\delta\phi_{<}^{\xi*}(\vec{k_{2}},\eta)
\rangle_{P}&=&\int_{\eta_{i}}^{\eta}d\eta_{1}\int_{\eta_{i}}^{\eta}d\eta_{2}\phi_{0}(\eta_{1})\phi_{0}
(\eta_{2})\\\nonumber&\times&
\langle{}\xi_{2}(\vec{k_{1}},\eta_{1})\xi_{2}^{*}(\vec{k_{2}},\eta_{2})\rangle_{P}\\
&\times&g(k_{1},\eta,\eta_{1})g(k_{2},\eta,\eta_{2}),
\end{eqnarray}
where\begin{eqnarray}\label{xi}\nonumber
&&\langle\xi_{2}(\vec{k_{1}},\eta_{1})\xi_{2}^{*}(\vec{k_{2}},\eta_{2})\rangle_{P}=
-288\lambda^{2}a^{4}(\eta_{1})a^{4}(\eta_{1})\\
&&\times(2\pi)^{3}\delta^{3}(\vec{k_{1}}-\vec{k_{2}})Re{}G_{++}^{\Lambda 2}(\eta_{1},\eta_{2},\vec{k_{1}}).
\end{eqnarray}
Thus $\Delta_{\phi^{\xi}}^{2}(k)$ can be expressed as
\begin{eqnarray}\label{espectro}
\Delta_{\phi^{\xi}}^{2}(k)&=&-\lambda^{2}\frac{144}{\pi^{2}}k^{3}\int_{\eta_{i}}^{\eta}d\eta_{1}
\int_{\eta_{i}}^{\eta}d\eta_{2}\;a^{4}(\eta_{1})a^{4}(\eta_{2})\nonumber\\
&\times&\phi_{0}(\eta_{1})\phi_{0}(\eta_{2})g(k,\eta,\eta_{1})g(k,\eta,\eta_{2})\nonumber\\
&\times&Re{}G_{++}^{\Lambda 2}(\eta_{1},\eta_{2},\vec{k}).
\end{eqnarray}
The Fourier transform $Re{}G_{++}^{\Lambda 2}(\eta_{1},\eta_{2},\vec{k})$
(where $k<\Lambda$) can be obtained from Eq.(\ref{ge++p}) (see Appendix \ref{AppendixA})
replacing $2{}k_0$ by $k$.

\begin{figure}
\begin{center}
\includegraphics[width=8cm]{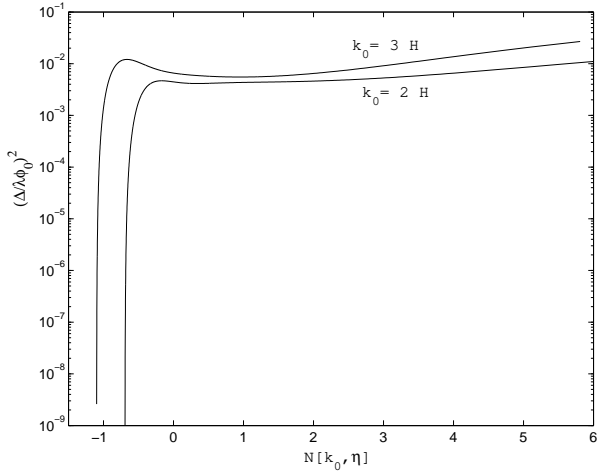}\\
\end{center}
\caption{$\Delta_{\phi^{\xi}}^{2}/(\lambda\phi_{0})^{2}$ as a
function of $\mathcal{N}[k_{0},\eta]$ for $\Lambda=5{}H$ and two different values of $k_{0}$. Note that the vertical axis is on a logarithmic scale.}
\label{potencia}
\end{figure}

Since the additional contribution $\Delta_{\phi^{\xi}}^{2}(k)$ to
the power spectrum is of order $\lambda^2$, it is expected to be
negligible. As we are assuming that $\phi_0$ is a slowly varying
field, we can compare the relative order of magnitude of both
contribution setting
$\phi_0(\eta_{1})\sim\phi_{0}(\eta_{2})\sim\phi_{0}$ and
$H^2\sim\lambda\phi_0^4/m_{\rm pl}^2$,  so that $\phi_{0}$ can be taken
out from the integration in Eq.(\ref{espectro}). Thereby, when
$|k\eta|<1$, the contributions  $\Delta_{\phi^{\xi}}^{2}(k)$ is
negligible compared to the usual $\Delta_{\phi^{q}}^{2}(k)$ if
$(\Delta_{\phi^{\xi}}(k)/\lambda\phi_0)^2<<\lambda^{-1}(\phi_0/m_{\rm pl})^2$.
Since the last quotient is usually much bigger than one, this
condition is typically satisfied if
$(\Delta_{\phi^{\xi}}(k)/\lambda\phi_0)^2<1$. As in the example
shown in Fig. \ref{potencia}, this is the case and hence the
additional contribution can be neglected.

\begin{figure}
\begin{center}
\includegraphics[width=8cm]{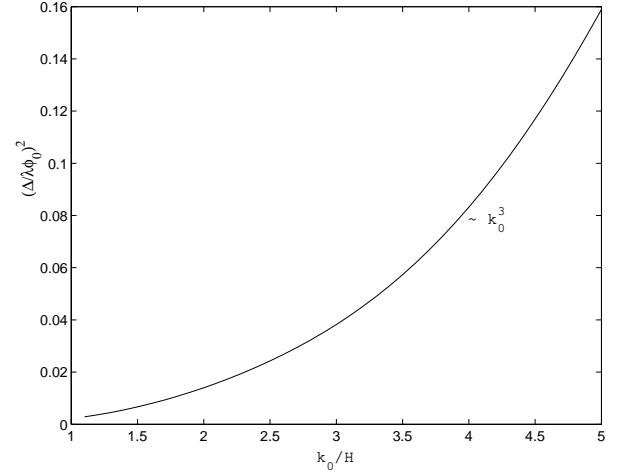}\\
\end{center}
\caption{$\Delta_{\phi^{\xi}}^{2}/(\lambda\phi_{0})^{2}$ as a
function of $k_{0}/H$ for $\Lambda=5{}H$ and
$|k_{0}\eta|=0.001$.}\label{potenciak}
\end{figure}
\begin{figure}
\begin{center}
\includegraphics[width=8cm]{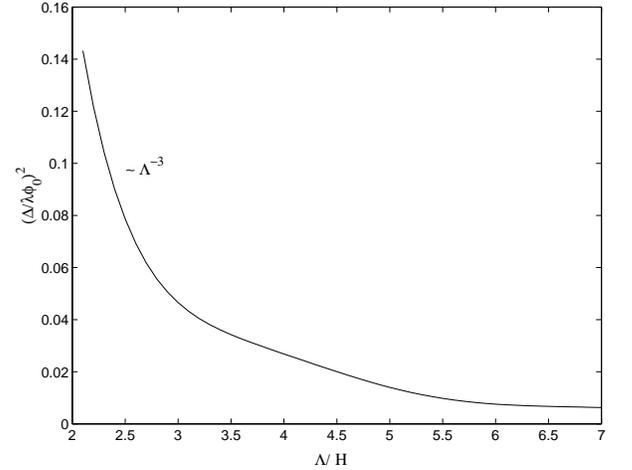}\\
\end{center}
\caption{$\Delta_{\phi^{\xi}}^{2}/(\lambda\phi_{0})^{2}$ as a
function of $\Lambda/H$ for $k_{0}=2{}H$ and
$|k_{0}\eta|=0.001$.}\label{potenciacut}
\end{figure}

On the other hand, the usual contribution
$\Delta_{\phi^{q}}^{2}(k)$ is independent of $k$ for a fixed value
of $k\eta$, corresponding to a nearly scale-invariant spectrum,
whereas $\Delta_{\phi^{\xi}}^{2}(k)$ depends on $k$ and $\Lambda$
(see Figs. \ref{potenciak} and \ref{potenciacut}). To see this
more clearly, it is useful to rewrite $\Delta_{\phi^{\xi}}^{2}(k)$
as
\begin{eqnarray}
\Delta_{\phi^{\xi}}^{2}(k)&=&-\frac{144}{\pi^{2}}
\int_{{k\eta_{i}}}^{k\eta}\frac{d{}x_{1}}{x_{1}^{4}}\int_{{k\eta_{i}}}
^{k\eta}\frac{d{}x_{2}}{x_{2}^{4}}\phi_{0}(\eta_{1})\phi_{0}(\eta_{2})\nonumber\\
&\times&f(k\eta,x_{1})f(k\eta,x_{2})F(x_{1},x_{2},{\frac{\Lambda}{k}}),
\end{eqnarray} where $f\equiv{}k^{3}H{}g$ and
$F\equiv{}k^{3}H^{-1}Re{G_{++}^{\Lambda 2}}$, which depend on $k$ not only
through the $k\eta$ combination, but also through the combinations
$k\eta_i$ and $\Lambda/k$. It is well-known that a finite duration
of the inflation stage produces a departure of the power spectrum
from the scale-invariant one \cite{Boya2004}. The breaking of the
scale invariance by the presence of the infrared cut-off $\Lambda$
is similar to the one found in Ref. \cite{CalzGoron}.

It is important to note that we have assumed a Gaussian initial
condition for the reduced density matrix elements and that the
Gaussian form of these elements is not affected by the
non-linearities of the interactions. However, as was pointed out
in Ref. \cite{diana}, it is expectable that when the
non-linearities become important, the Gaussian approximation
breaks down and therefore the associated Wigner functional becomes
non-positive. Nevertheless, as well as for a given mode the
Gaussian approximation remains valid up to times longer than the
decoherence time-scale, one can use the classical description for
it even in the non-linear regime.

\section{Final remarks}

Let us summarize the results contained in this paper. After the
integration of the high frequency modes in Section II, we obtained
the CGEA for the modes whose wave vector is shorter than a
critical value $\Lambda$. From the imaginary part of the CGEA we
obtained, in Section III, the diffusion coefficients of the master
equation. System and environment are two sectors of a single
scalar field, and the results depend on the  ``size" of these
sectors, which is fixed by the cut-off $\Lambda$.

To analyze the decoherence process, in Section IV we evaluated the
diffusion coefficients and its integrations over the conformal
time. It allows us to conclude that if we consider a cut-off
$\Lambda \sim  H$, those modes with wave vector $k_0 \ll
\Lambda$ are the more affected by diffusion through one of the
coefficients.
For these modes, we saw that the effect is weakly
dependent of the critical wave vector $\Lambda$. If one consider a
cut-off $\Lambda \geq H$, and modes $H < k_0 < \Lambda$, diffusive
effects are larger for those modes in the system whose wave vector
is close to the critical $\Lambda$ \cite{lombmazz}.

We presented the complete expression of the diffusive terms (in
Appendix A), and also some simple analytical approximations to the
coefficients, which are useful to make an evaluation of
the decoherence time-scale. We performed an extensive analysis of
the evaluation of the time-scale for the decoherence process for a
typical case of interest. In such a case, we obtained that for a
given mode $k_0 < \Lambda \lesssim H$, decoherence is effective by
the time in which inflation is ending.

In Section V we analyzed the effective evolution of the system.
Assuming an homogeneous system-field we first presented the explicit form of the
renormalized stochastic Langevin equation and we analyzed the 
relative importance of the terms appearing in that equation due to the 
system-environment interaction. From such analysis
we conclude that those terms are of the same order of magnitude than the  
classical ones for times longer than the typical decoherence time-scale.

We them considered inhomogeneity generation in a simple particular
situation, in which we analyzed the influence of the environment
on the power spectrum for some modes in the system. In that
situation, splitting the spectrum as the sum of the contribution
coming from the unitary evolution of the Bunch-Davies initial
condition plus the one which appears because of the
system-environment interaction, we found that the latter is
negligible compared to the former. In spite of this result, we
have remarked that the system-environment interaction is essential
to have a complete quantum to classical transition which allows
a late-time classical treatment of the system degrees of freedom.

As for future work, we consider that it is worth applying the same
procedure used in this paper to inflationary models involving two
or more interacting fields, such as some hybrid model
\cite{Linde1994}.

%
\section{Acknowledgments}
This work is supported by UBA, CONICET and Fundaci\'on Antorchas;
Argentina. We specially thank M. Zaldarriaga for very useful comments and 
suggestions. We also thank comments from E. Calzetta, B.L. Hu, C. Kiefer,
F. Mazzitelli, N. Mavromatos, and R. Rivers.
\appendix

\section{Diffusion Coefficients}\label{AppendixA}
In this appendix we describe some
technical details about the computation of the diffusion coefficients
${D}_{1}$ and ${D}_{2}$, starting from Eq.(\ref{D1}) and (\ref{D2})
respectively.

In order to evaluate the coefficient ${D}_{1}$, we have to perform
the Fourier transform of the imaginary part of the propagator for
the environment $G_{++}^{\Lambda}$ (see Eq.(\ref{D1})). This
propagator can be expressed in terms of the mode functions
$\phi_k$ given by
\begin{eqnarray}\label{modoent}\nonumber
\phi_{\vec{k}}&=&\frac{1}{(2\pi)^{3/2}}\frac{\exp\{{-ik\eta}\}}{\sqrt{2k}a(\eta)}\left(1-\frac{i}{k\eta}\right)\exp\{i\vec{k}\cdot\vec{x}\}\\
&\equiv&\tilde{\phi}_{k}(\eta)\exp\{i\vec{k}\cdot\vec{x}\},
\end{eqnarray} which corresponds to the  Bunch-Davies vacuum assumed for the
environment-field. The imaginary part of the propagator reads
\begin{equation}\label{img++}
Im{}G^{\Lambda}_{++}(x_1,x_2)=\int_{k>\Lambda}d^{3}k\;Re\{\phi_{\vec{k}}(x_{1})\phi^{*}_{\vec{k}}(x_{2})\}.
\end{equation}
Substituting this expression into  Eq.(\ref{D1}) we obtain
\begin{eqnarray}\label{coefD1}\nonumber
{D}_{1}(\vec{k_{0}},\eta,\Lambda)&=&\frac{\tilde{k_{0}^{4}}}{2} \int_{x}^{\tilde{{k_{0}}}} dy\left[\frac{\cos[3\Delta_{xy}]}{{6}x^{3}y^{3}}\left(1+\frac{1}{{9}{x}y}\right)\right.\;\;\;\;\;\;\\
&+&\left.\frac{\sin[3\Delta_{xy}]\Delta_{xy}}{{18}x^{4}y^{4}}\right]{F}^{3}_{cl}(x,y)\;\Theta(3-\tilde{\Lambda}),\;
\end{eqnarray} with  \begin{equation}\label{Fcladim}
{F}_{cl}(x,y)=\frac{\sin[\Delta_{xy}]}{x}+\frac{y\cos[\Delta_{xy}]}{x},
\end{equation}where to reduce the notation we have defined the following set of dimensionless variables:
\begin{eqnarray}\label{varadim}\nonumber
\tilde{k_{0}}=\frac{k_{0}}{H}\;,\;
\tilde{\Lambda}=\frac{\Lambda}{k_{0}}\;,\;
y=|k_{0}\eta'|\;,\;x=|k_{0}\eta|\;,\;\Delta_{xy}=x-y.
\end{eqnarray}
The integrations above can be performed exactly and the result can
be numerically integrated  over the conformal time $\eta$.

To obtain ${D}_{2}$ from Eq.(\ref{D2}) we need to compute the
Fourier transform of the real part of ${G}^{\Lambda{}2}_{++}$ in
$\vec{k}=2\vec{k_{0}}$ and $\vec{k}=\vec{0}$. This real part is
given by
\begin{eqnarray}\label{Rege++}\nonumber
{Re\;G}^{\Lambda{}2}_{++}(\eta,\eta',2\vec{k_{0}})=-(2\pi)^{3}\int_{k>\Lambda}d^{3}k\int_{k'>\Lambda}d^{3}k'\;\;&&\\
\times\delta^{3}\left(\vec{k}+\vec{k}'-2\vec{k_{0}}\right)Re\{\tilde{\phi}_{k}(\eta)\tilde{\phi}^{*}_{k}(\eta')\tilde{\phi}_{k'}(\eta)\tilde{\phi}^{*}_{k'}(\eta')\},&&
\end{eqnarray}where $\tilde{\phi}_{k}$ is the time-dependent mode function defined in Eq.(\ref{modoent}) and  $\tilde{\phi}^{*}_{k}$ its complex conjugate.
This Fourier transform can be expressed in terms of
integrals of the form,
\begin{eqnarray}\label{inm}\nonumber
I_{n,m}^{C}&=&\frac{k_{0}^{m+n-3}}{\pi}\int_{k>\Lambda}{d^{3}k}\int_{k'>\Lambda}{d^{3}k'}\delta^{3}\left(\vec{k}+\vec{k}'-2\vec{k_{0}}\right)\\\nonumber
&\times&\frac{\cos[(k+k')(\eta-\eta')]}{k^{n}{k'}^{m}}\\ \nonumber
&=&\int_{\tilde{\Lambda}+2}^{+\infty}\frac{d{u}}{u^{n-1}}\int_{u-2}^{u+2}\frac{d{z}}{z^{m-1}}\cos[(u+z)\Delta_{xy}]\\
&+&\int_{\tilde{\Lambda}}^{\tilde{\Lambda}+2}\frac{d{u}}{u^{n-1}}\int_{\tilde{\Lambda}}^{u+2}\frac{d{z}}{z^{m-1}}\cos[(u+z)\Delta_{xy}],\;
\end{eqnarray}where $n$ and $m$ are integer numbers (only $m=3$ result to be necessary).
The second equality follows after the change of variables
$u=k_{0}^{-1}k$ and
$z=k_{0}^{-1}|\vec{k}-2\vec{k_{0}}|=k_{0}^{-1}\sqrt{k^{2}-4{}k{}k_{0}\cos(\theta)+4{}k_{0}^{2}}$,
where $\theta$ is the angle between $\vec{k}$ and $\vec{k_{0}}$.
We also define the integral $I_{n,m}^{S}$ as the one obtained
from $I_{n,m}^{C}$ by replacing the cosine function by sine.

An integration of $I_{n,m}^{C,S}$ by parts  yields
\begin{eqnarray}\label{inte}\nonumber
I_{n,m}^{C}+\Delta_{xy}\frac{I_{n,m-1}^{S}}{2-m}&=&\int_{\tilde{\Lambda}}^{+\infty}\frac{d{u}}{u^{n-1}}\frac{\cos[2(u+1)\Delta_{xy}]}{(2-m)(u+2)^{m-2}}\;\;\;\;\;\\
\nonumber
&-&\int_{\tilde{\Lambda}+2}^{+\infty}\frac{d{u}}{u^{n-1}}\frac{\cos[2(u-1)\Delta_{xy}]}{(2-m)(\tilde{\Lambda}+u)^{m-2}}\;\;\;\;\\
&-&\int_{\tilde{\Lambda}}^{\tilde{\Lambda}+2}\frac{d{u}}{u^{n-1}}\frac{\cos[2(u+1)\Delta_{xy}]}{(2-m){\tilde{\Lambda}}^{m-2}},
\end{eqnarray}\begin{eqnarray}\label{inte}
\nonumber
I_{n,m}^{S}-\Delta_{xy}\frac{I_{n,m-1}^{C}}{2-m}&=&-\int_{\tilde{\Lambda}}^{+\infty}\frac{d{u}}{u^{n-1}}\frac{\sin[2(u+1)\Delta_{xy}]}{(2-m)(u+2)^{m-2}}\;\;\\
\nonumber
&+&\int_{\tilde{\Lambda}+2}^{+\infty}\frac{d{u}}{u^{n-1}}\frac{\sin[2(u-1)\Delta_{xy}]}{(2-m)(\tilde{\Lambda}+u)^{m-2}}\;\;\;\;\\
&+&\int_{\tilde{\Lambda}}^{\tilde{\Lambda}+2}\frac{d{u}}{u^{n-1}}\frac{\sin[2(u+1)\Delta_{xy}]}{(2-m){\tilde{\Lambda}}^{m-2}}.
\end{eqnarray}With the use of these properties and definitions, the Fourier transform (\ref{Rege++}) becomes
\begin{eqnarray}\label{ge++p}
{ReG}^{\Lambda2}_{++}(\eta,\eta',2\vec{k_{0}})=\frac{-H^{4}x^{2}y^{2}}{8(2\pi)^{2}k_{0}^{3}}\left\{I_{A}+\frac{2I_{B}}{x{}y}+\frac{I_{C}}{x^{2}{y}^{2}}\right\}\;\;\;
\end{eqnarray}where we have defined:\begin{subequations}\label{inte}
\begin{align}I_{A}=&I_{11}^{C}\\
I_{B}=&I_{31}^{C}-\Delta_{xy}I_{21}^{S}\\
I_{C}=&I_{33}^{C}-2\Delta_{xy}I_{32}^{S}-\Delta_{xy}^2{}I_{22}^{C}
\end{align}\end{subequations}
These last integrals are easily computed,  with the
result\begin{eqnarray}
I_{A}&=&2\pi\delta\left(\Delta_{xy}\right)+\frac{\cos[2(\tilde{\Lambda}+1)\Delta_{xy}]-\cos[2\tilde{\Lambda}\Delta_{xy}]}{\Delta_{xy}^2},\;\;\\\nonumber
I_{B}&=&\cos[2\Delta_{xy}]\left\{\mathcal{C}i[2(\tilde{\Lambda}+2)|\Delta_{xy}|]-\mathcal{C}i[2\tilde{\Lambda}|\Delta_{xy}|]\right\}\\
\nonumber&-&\sin[2|\Delta_{xy}|]\left\{\pi-\mathcal{S}i[2(\tilde{\Lambda}+2)|\Delta_{xy}|]-\mathcal{S}i[2\tilde{\Lambda}|\Delta_{xy}|]\right\}\\
&+&\frac{\sin[2(\tilde{\Lambda}+1)\Delta_{xy}]}{\tilde{\Lambda}\Delta_{xy}}-\frac{\sin[2\tilde{\Lambda}\Delta_{xy}]}{\tilde{\Lambda}\Delta_{xy}},
\end{eqnarray}

\begin{eqnarray}\nonumber
I_{C}&=&\left\{\pi-\mathcal{S}i[2(\tilde{\Lambda}+2)|\Delta_{xy}|]-\mathcal{S}i[2\tilde{\Lambda}|\Delta_{xy}|]\right\}\\\nonumber
\nonumber&\times&\left(\cos[2\Delta_{xy}]|\Delta_{xy}|-\frac{\sin[2|\Delta_{xy}|]}{2}\right)\\
\nonumber&+&\left\{\mathcal{C}i[2(\tilde{\Lambda}+2)|\Delta_{xy}|]-\mathcal{C}i[2\tilde{\Lambda}|\Delta_{xy}|]\right\}\\
\nonumber&\times&\left(\sin[2\Delta_{xy}]\Delta_{xy}+\frac{\cos[2\Delta_{xy}]}{2}\right)\\
&+&\frac{\cos[2\tilde{\Lambda}\Delta_{xy}]}{\tilde{\Lambda}^{2}}-\frac{\cos[2(\tilde{\Lambda}+1)\Delta_{xy}]}{\tilde{\Lambda}},
\end{eqnarray}where $\mathcal{S}i[x]$ and $\mathcal{C}i[x]$ are the sine and the cosine integral, respectively \cite{Abramob}.

The Fourier transform in $\vec{k}=\vec{0}$ is much simpler to
compute, and the result is
\begin{eqnarray}\label{FTinzero}\nonumber
&&{Re{}G}^{\Lambda{}2}_{++}(\eta,\eta',\vec{0})=\frac{-H^{4}x^{2}y^{2}}{2(2\pi)^{2}k_{0}^{3}}\left\{
\frac{\pi}{2}\delta\left(\Delta_{xy}\right)\right.\;\;\;\;\;\;\;\;\;\;\;\\\nonumber
&&\;\;\;\;-\;\sin[2\tilde{\Lambda}\Delta_{xy}]\left(\frac{1}{2\Delta_{xy}}+\frac{2}{3}\frac{\Delta_{xy}}{\tilde{\Lambda}^{2}x^{2}y^{2}}\right)\\
 \nonumber&&\;\;\;\;+\;\frac{\cos[2\tilde{\Lambda}\Delta_{xy}]}{\tilde{\Lambda}}\left(\frac{2}{x{}y}+\frac{\Delta_{xy}^{2}}{3{}x^{2}y^2}+\frac{1}{3\tilde{\Lambda}^{2}x^{2}y^2}\right)\\
&&\;\;\;\;\left.-\;(\pi-2\mathcal{S}i[2\tilde{\Lambda}|\Delta_{xy}|])\frac{|\Delta_{xy}|}{x{}y}\left(\frac{\Delta_{xy}^{2}}{3{}x{}y}+1\right)\right\}.\;
\end{eqnarray}

Substitution of Eqs.(\ref{ge++p}) and (\ref{FTinzero}) into
Eq.(\ref{D2}) yields an expression for ${D}_{2}$ that allows  to
evaluate it and its integration over the conformal time $\eta$
numerically.

\section{Renormalization}\label{AppendixB}

In this Appendix we obtain the renormalized semiclassical equation
for the system-field given in Eq.(\ref{ERen}). To do so, we
follow the same procedure as for the renormalization of the
evolution equation for the mean value $\langle\hat{\phi}\rangle$
\cite{RenoPazMazziCarmen}.

Starting with the bare action for the field $\phi$, the semiclassical equation for the homogeneous system-field $\phi_<$ reads
\begin{eqnarray} \label{ecSClasDSaren}\nonumber
&&\phi_{<}''(\eta)+[m_{0}^{2}+\xi_{0}R]a^{2}(\eta)\phi_{<}(\eta)+2\mathcal{H}\phi'_{<}(\eta)\\\nonumber
&&+4{}\lambda_{0}{}a^{2}(\eta){}\phi^{3}_{<}(\eta)-12\lambda{}a^{2}(\eta)\phi_{<}(\eta){i}G^{\Lambda}_{++}(\eta,\eta,\vec{0})\\\nonumber
&&-288\lambda^{2}{a}^{2}(\eta)\phi_{<}(\eta)\int_{\eta_{i}}^{\eta}d\eta'{}a^{4}(\eta')\int{}d^{3}y\phi^{2}_{<}(\eta'){}\\
&&\times{}ImG^{\Lambda 2}_{++}(\eta,\eta',\vec{x}-\vec{y})=-\xi_{2}(\eta,\vec{x})\frac{\phi_{<}(\eta)}{a^{2}(\eta)},
\end{eqnarray} where $m_{0}$, $\xi_{0}$ and $\lambda_{0}$ are the bare constants, and  $R=12{H^2}$. Here we have replaced $\lambda_{0}$ by the renormalized constant $\lambda$ in the one-loop terms.

Let us add and substrate the following terms to the  left-hand side of Eq.(\ref{ecSClasDSaren}):
 \begin{equation}\label{terms}
 12\lambda{}a^2(\eta)\phi_<(\eta)\langle\hat{\phi}_>^{2}\rangle_{ad2}
 \end{equation} where $\langle\hat{\phi}_>^{2}\rangle_{ad2}$ is the adiabatic expansion  of the
 expectation value $\langle\hat{\phi}_>^{2}\rangle$  up to the second
 adiabatic order \cite{RenoPazMazziCarmen}.

In what follows we will show how the infinities in Eq.(\ref{ecSClasDSaren}) are cancelled by the terms
subtracted. We will compute the divergent terms added via dimensional regularization,
which will thus be able to be absorbed in the bare parameters as usual.

The expansion of $\langle\hat{\phi}_>^{2}\rangle$ up to the second
adiabatic order yields:
\begin{equation}
\langle\phi_>^{2}\rangle_{ad2}\equiv\langle\phi_>^{2}\rangle_{ad2}^{F}+\langle\phi_>^{2}\rangle_{ad2}^{I},
\end{equation}
with
\begin{subequations}\label{ad}\begin{align}
\langle\phi_>^{2}\rangle_{ad2}^{F}=&\frac{a}{2}\int_{k>\Lambda}\frac{d^{3}k}{(2\pi{}a)^{3}}\left\{\frac{1}{\omega_{k}}-\frac{\left(\xi_{0}-\frac{1}{6}\right)a^{2}R}{2\omega_{k}^{3}}\right.\\
\nonumber
+&\left.\frac{m_{0}^{2}}{4\omega_{k}^{5}}[{a'}^{2}+a{}a'']-\frac{5{}m_{0}^{4}}{8\omega_{k}^{7}}(a{}a')^{2}\right\},\\\label{ad2}
\langle\phi_>^{2}\rangle_{ad2}^{I}=&-\frac{a}{2}\int_{k>\Lambda}\frac{d^{3}k}{(2\pi{}a)^{3}}\frac{6\lambda_{0}{}a^{2}}{\omega_{k}^{3}}\phi_{<}^{2}.
\end{align}\end{subequations} where $\omega_{k}^2=k^2+a^2 m_0^2$, and the integrations over the wave vector $\vec{k}$ are restricted by $k>\Lambda$.

 In order to use dimensional regularization, we perform the integrations above over all wave
 vectors with $k \geq 0$  and then we subtract the ones restricted by $k<\Lambda$ with the result
\begin{subequations}\label{adreg}
 \begin{align}\nonumber
 \langle\phi^{2}_>\rangle_{ad2}^{F}&=-\frac{1}{4\pi^{2}}\frac{R}{12}\left[\frac{1}{n-4}+\frac{\gamma}{2}-\ln\Big{|}\frac{a(\eta)\mu}{2\Lambda}\Big{|}\right]\\
 &-\frac{\Lambda^{2}}{8\pi^{2}a^{2}(\eta)},\\\label{adreg2}
\langle\phi^{2}_>\rangle_{ad2}^{I}&=\frac{3\lambda}{2\pi^{2}}\phi_{<}^{2}(\eta)\left[\frac{1}{n-4}+\frac{\gamma}{2}-\ln\Big{|}\frac{a(\eta)\mu}{2\Lambda}\Big{|}\right],
\end{align}
\end{subequations} where $\gamma$ is the Euler's constant and we have replaced the bare parameters with their renormalized counterparts ($m=0$, $\xi=0$ and $\lambda$).
It is important to  notice that only after subtracting the bare
parameters can be replaced by the renormalized ones, since one of
the integrals over $k<\Lambda$ has an infrared logarithmic
divergence.

Writing the bare parameters in terms of the renormalized ones plus
conterterms,
\begin{equation}\nonumber
m_{0}^{2}=0+\delta{}m^{2}\;\;,\;\;
\xi_{0}=0+\delta\xi\;\;,\;\; \lambda_{0}=  \lambda+
\delta\lambda,
\end{equation} we can see that the divergences appearing for
 $n\rightarrow{}4$ are cancelled with the use of the following counterterms:
\begin{subequations}\label{conter} \begin{align}
 \delta{}m^{2}=&\Delta{}m^{2},\\
\delta{}\xi=&\frac{\lambda}{4\pi^{2}}\left[\frac{1}{n-4}\right]+\Delta{}\xi,\\
\delta{}\lambda=&-\frac{9\lambda^{2}}{2\pi^{2}}\left[\frac{1}{n-4}\right]+\Delta{}\lambda,
\end{align}
\end{subequations}where $\Delta{}m^{2}$, $\Delta{}\xi$ and
$\Delta{}\lambda$ remain finite as $n\rightarrow{}4$.

Replacing the bare parameters by the renormalized ones in the integrals over $k>\Lambda$ of  Eq.(\ref{ad}) we obtain
\begin{subequations}\label{terminos} \begin{align}
\langle\phi^{2}_>\rangle_{ad2}^{F}&=-i{}G_{++}^{\Lambda}(\eta,\eta,\vec{0}),\\
\langle\phi^{2}_>\rangle_{ad2}^{I}&=-\frac{3\lambda}{2\pi^{2}}\int_{k>\Lambda}\frac{d{}k}{k}\phi_{<}^{2}(\eta).
\end{align}\end{subequations}From equation above it is simple to note that the order $\lambda$
contribution  in Eq.(\ref{ecSClasDSaren}) is completely cancelled by the
one with $\langle\phi^{2}_>\rangle_{ad2}^{F}$ (see Eq.(\ref{terms})). 

In order to separate the divergent part from the order $\lambda^2$ contribution of
Eq.(\ref{ecSClasDSaren}), we write the propagator as
\begin{eqnarray}\label{imagpop}\nonumber
   \int{}d^{3}yImG^{\Lambda 2}_{++}(\eta,\eta',\vec{x},\vec{y})=-(2\pi)^{3}
   \int_{k>\Lambda}d^{3}k\;Im[\tilde{\phi}_{\vec{k}}^{2}\;\tilde{\phi}^{*2}_{\vec{k}}]\;\;&&\\\nonumber
      =-\int_{k>\Lambda}\frac{d{}k}{8\pi^{2}}\left\{\frac{\cos[2{}k(\eta-\eta')]}{a^{2}(\eta
      )a^{2}
   (\eta')}\left[\frac{2(\eta-\eta')}{k\eta\eta'}+\frac{2(\eta-\eta')}{k^{3}\eta^{2}{\eta'}^{2
   }}\right]\right.\;\;&&\\
\nonumber
-\left.\frac{\sin[2{}k(\eta-\eta')]}{a^{2}(\eta)a^{2}(\eta')}\left[1+\frac{1}{k^{4}\eta^{2}{\eta'}^{2}}+\frac{2}{k^{2}\eta\eta'}-\frac{(\eta-\eta')^{2}}{k^{2}\eta^{2}{\eta'}^{2}}\right]\right\}&&\\\nonumber
\equiv{}
I_{D}+I_{ND},\;\;\;\;\;\;\;\;\;\;\;\;\;\;\;\;\;\;\;\;\;\;\;\;\;\;\;\;\;\;\;\;\;\;\;\;\;\;\;\;\;\;\;\;\;\;\;\;\;\;\;\;\;\;\;\;\;\;\;\;\;\;\;\;\;\;\;\;\;&&
\end{eqnarray}where  $\tilde{\phi}_{\vec{k}}$ and $\tilde{\phi}^{*}_{\vec{k}}$ are the  mode function and its complex conjugate respectively, given in Eq.(\ref{modoent}). The only divergent term is
\begin{equation}\label{Id}
I_{D}\equiv\int_{k>\Lambda}\frac{d{}k}{8\pi^{2}}\frac{\sin[2{}k(\eta-\eta')]}{a^{2}(\eta)a^{2}(\eta')}.
\end{equation} With the use of the definition of  $I_{D}$ and $I_{ND}$, we can write the order $\lambda^2$ contribution as
\begin{eqnarray}
&-&288\;\lambda^{2}{a}^{2}(\eta)\int_{\eta_{i}}^{\eta}d\eta'{}a^{4}(\eta'){}\phi_{<}(\eta)\int{}d^{3}y\;\phi^{2}_{<}(\eta'){}ImG^{\Lambda 2}_{++}\nonumber\\&\equiv&
12\lambda\;a^{2}(\eta)\phi_{<}(\eta)\left([\langle\phi^{2}_{>}\rangle]^{I}_{D}+[\langle\phi^{2}_{>}\rangle]^{I}_{ND}\right),
\end{eqnarray}with\begin{eqnarray}\label{phiD}\nonumber
&&[\langle\phi^{2}_{>}\rangle]^{I}_{D}=-24\lambda\int_{\eta_{i}}^{\eta}d\eta'\;a^{4}(\eta')\phi_{<}^{2}(\eta')I_{D}\\
\nonumber
&&=-\frac{3\lambda}{\pi^{2}}\int_{\eta_{i}}^{\eta}d\eta'\frac{a^{2}(\eta')\phi_{<}^{2}(\eta')}{a^{2}(\eta)}\int_{k>\Lambda}d{}k\;\sin[2{}k(\eta-\eta')]\\
\nonumber
&&=-\frac{3\lambda}{2\pi^{2}}\phi_{<}^{2}(\eta)\int_{k>\Lambda}\frac{d{}k}{k}\;+\;\frac{3\lambda}{2\pi^{2}}\frac{a^{2}(\eta_{i})\phi_{<}^{2}(\eta_{i})}{a^{2}(\eta)}\\
\nonumber
&&\times\int_{k>\Lambda}\frac{d{}k}{k}\cos[2{}k(\eta-\eta_{i})]\\\nonumber
&&+\frac{3\lambda}{2\pi^{2}}\int_{\eta_{i}}^{\eta}d\eta'\frac{(a^{2}(\eta')\phi_{<}^{2}(\eta'))'}{a^{2}(\eta)} \int_{k>\Lambda}\frac{d{}k}{k}\cos[2{}k(\eta-\eta')],
\end{eqnarray}where the last equality follows after performing an integration by parts.
In this expression we can see that the first term after the last equality is
$\langle\phi^{2}_>\rangle_{ad2}^{I}$ and hence all infinities are
cancelled. Finally, performing the $I_{ND}$ integral and reordering the terms we obtain the explicit form of the renormalized equation given in Eq.(\ref{ERen}).

%
%

\bibliography{apssamp}

\begin{thebibliography}{00}

\bibitem{linde}A.D. Linde, {\it Particle Physics and Inflationary Cosmology}, Harwood, Chur,
Switzerland (1990)

\bibitem{Boya2004} D. Boyanovsky, H.J. de Vega, and N.G. Sanchez,{\it The Classical 
and Quantum Inflaton: the Precise Inflationary Potential and Quantum Inflaton Decay after WMAP}
[astro-ph/0503128]

\bibitem{todosLSS}S.W. Hawking, Phys. Lett. {\bf B115}, 295 (1982); A.A. Starobinsky,
Phys. Lett. {\bf B117}, 175 (1982); A.H. Guth and S.Y. Pi, Phys.
Rev. Lett. {\bf 49}, 1110 (1982)

\bibitem{old} D.A. Kirzhnits and A.D. Linde, Phys. Lett. {\bf B42}, 471 (1972)

\bibitem{cormier} S.A. Ramsey and B.L. Hu, Phys. Rev. {\bf D56}, 678 (1997);
S.A. Ramsey, B.L. Hu, and A.M. Stylianopoulos,  Phys. Rev. {\bf
D57}, 6003 (1998); D. Cormier and R. Holman, Phys. Rev. {\bf D62},
023520 (2000) and references therein

\bibitem{Kibble} T.W.B. Kibble, Phys. Rep. {\bf 67}, 183 (1980)

\bibitem{vilen} A. Vilenkin, Phys.Rep. {\bf 121}, 263 (1985);  A. Rajantie, Int. J.
Mod. Phys. {\bf A17}, 1 (2002)


\bibitem{lomplb}  F.C. Lombardo, F.D. Mazzitelli, and R.J. Rivers, Phys. Lett.
{\bf B523}, 317 (2001)

\bibitem{deconpb}  F.C. Lombardo, F.D. Mazzitelli, and R.J. Rivers, Nucl. Phys.
{\bf B672}, 462 (2003)

\bibitem{diana} F.C. Lombardo, F.D. Mazzitelli, and D. Monteoliva, Phys. Rev.
{\bf D62}, 045016 (2000)

\bibitem{lomplb2}  R.J. Rivers, F.C. Lombardo, and F.D. Mazzitelli, Phys.
Lett. {\bf B539}, 1 (2002)

\bibitem{calzettahu95} E. Calzetta and B.L. Hu, Phys. Rev. {\bf D52}, 6770 (1995)

\bibitem{guthpi} A. Guth and S.Y. Pi, Phys. Rev. {\bf D32}, 1899 (1985)

\bibitem{staro}D. Polarski and A.A. Starobinsky, Class. Quantum Grav. {\bf 13}, 377 (1996);
J. Lesgourgues, D. Polarski and A.A. Starobinsky, Nucl. Phys. {\bf
B497}, 479 (1997)

\bibitem{Mopntemayor} M. Bellini, H. Casini, R. Montemayor, and P.Sisterna, Phys. Rev. 
{\bf D54}, 7172 (1996); H. Casini, R. Montemayor, and P.Sisterna  Phys. Rev.  {\bf D59}, 063512 (1999)

\bibitem{kiefer}C. Kiefer, D. Polarski, and A.A. Starobinsky, Int. J. Mod. Phys. {\bf D7}, 455 (1998)

\bibitem{halli}J.J. Halliwell, Phys. Rev. {\bf D36}, 3626 (1987)

\bibitem{stancioff}F. Cooper, S.Y. Pi, and P. Stancioff, Phys. Rev. {\bf D34}, 3831 (1986)

\bibitem{giulinibook}D. Giulini, C. Kiefer, E. Joos, J. Kupsch, I.O. Stamatescu, and H.D.
Zeh, {\it Decoherence and the apperance of a classical world in
quantum theory},  Springer, Berlin, Germany (1996)

\bibitem{nuno} N.D. Antunes, F.C. Lombardo and D.  Monteoliva, Phys. Rev.
{\bf E64}, 066118 (2001)

\bibitem{matacz} A. Matacz, Phys. Rev. {\bf D55}, 1860 (1997)

\bibitem{Riotto}Sabino Matarrese, Marcello A.Musso, and Antonio Riotto,
JCAP {\bf 0405} 008 (2004); Michele Liguori,  Sabino Matarrese,
Marcello A.Musso, and Antonio Riotto, JCAP  {\bf 0408} 011 (2004)

\bibitem{HUPAZYHANG}  B. L. Hu, J. P. Paz and Y. Zhang, {\it Quantum Origin of Noise and 
Fluctuations in Cosmology}, in The Origin of Structure in the Universe, edited by E. Gunzig 
and P. Nardone (Kluwer, Dordrecht, 1993), p. 227.

\bibitem{lombmazz} F.C. Lombardo and F.D. Mazzitelli, Phys. Rev. {\bf D53}, 2001 (1996)

\bibitem{BirreLLDavies}N.D.Birrell and P.C.Davies, {\it Quantum fields in curved space},
Cambridge University Press (1982)

\bibitem{unruh} W. G. Unruh and W. H. Zurek Phys. Rev. {\bf D40}, 1071 (1989); A. Caldeira and A.
Leggett, Phys. Rev. {\bf A31}, 1059 (1985)

\bibitem{qbm}  B.L. Hu, J.P. Paz, and Y.
Zhang, Phys. Rev. {\bf D45}, 2843 (1992); {\bf D47}, 1576 (1993);
J.P. Paz, S. Habib, and W.H. Zurek, Phys. Rev. {\bf D47}, 488
(1993)

\bibitem{paula} F.C. Lombardo and P.I. Villar, Phys. Lett. {\bf A336}, 16 (2005)



\bibitem{langlois} David Langlois, Lectures delivered at the Cargese School of Physics and Cosmology, Cargese, France, August 2003
[hep-th/0405053 v1]

\bibitem{Peacock} John A. Peacock, {\it Cosmological Physics} (Cambridge University
Press, 1999); Scott Dodelson {\it Modern Cosmology} (Academic Press, 2003)

\bibitem{GMuler}C. Greiner and B. M\"{u}ler, Phys. Rev. {\bf D55}, 1026 (1997)

\bibitem{BLHU} B. L. Hu, {\it Quantum Statistical Field Theory in Gravitation an Cosmology}, in Proc. 
Third International Workshop on Thermal Field Theories and Applications, eds.
R. Kobes and G. Kunstatter (World Scientific, Singapore, 1994)
[gr-qc/9403061 v1]

\bibitem{RenoPazMazziCarmen}J.P. Paz and F.D. Mazzitelli,  Phys. Rev. {\bf D37}, 2170 (1988);
C. Molina-Par\'\i s and P. R. Anderson and S. A. Ramsey, Phys. Rev. {\bf D61},
127501 (2000)

\bibitem{Abramob}M. Abramowitz and I. Stegun, {\it Handbook of Mathematical Functions} 
(Dover Publications (N.Y.), 1972)

\bibitem{Morika} Hiroto  Kubotani and Tomoko Uesugi and Masahiro  Morikawa and Akio Sugamoto,
Prog. of Theor. Phys., {\bf 98}, 1063 (1997)

\bibitem{CalzGoron} Esteban A. Calzetta and Sonia Goronazky, Phys. Rev.  {\bf D55}, 1812 (1997)

\bibitem{Linde1994} Andrei Linde, Phys. Rev. {\bf D49}, 748 (1994)
\end{thebibliography}

\end{document}